\documentclass{article}

\usepackage{microtype}
\usepackage{graphicx}
\usepackage{subfigure}
\usepackage{booktabs} % for professional tables
\usepackage{multirow}
\usepackage{tikz}
\usepackage{bbding}
\usepackage{siunitx}
\usepackage{xcolor,colortbl}

\usepackage{hyperref}

\newcommand{\modelname}{MELP}

\newcommand{\para}[1]{\vspace{.05in}\noindent\textbf{#1}}
\newcommand{\highlightgain}[1]{\textbf{\textcolor{darkgreen}{#1}}}
\usepackage[accepted]{icml2025}

\usepackage{amsmath}
\usepackage{amssymb}
\usepackage{mathtools}
\usepackage{amsthm}

\usepackage{tabu}
\usepackage{xcolor}

\usepackage[capitalize,noabbrev]{cleveref}

\theoremstyle{plain}

\theoremstyle{definition}

\theoremstyle{remark}

\usepackage[textsize=tiny]{todonotes}

\icmltitlerunning{From Token to Rhythm: A Multi-Scale Approach for ECG-Language Pretraining}

\begin{document}

\twocolumn[
\icmltitle{From Token to Rhythm: A Multi-Scale Approach \\ for ECG-Language Pretraining}



\icmlsetsymbol{equal}{*}

\begin{icmlauthorlist}
\icmlauthor{Fuying Wang}{equal,hku}
\icmlauthor{Jiacheng Xu}{equal,hku}
\icmlauthor{Lequan Yu}{hku}
\end{icmlauthorlist}

\icmlaffiliation{hku}{School of Computing and Data Science, The University of Hong Kong, Hong Kong SAR, China}

\icmlcorrespondingauthor{Lequan Yu}{lqyu@hku.hk}

\icmlkeywords{ECG, Multimodal Foundation Model, AI for healthcare}

\vskip 0.3in
]



\printAffiliationsAndNotice{\icmlEqualContribution} 
% otherwise use the standard text.

% Training deep learning models to analyze heart signals (ECGs) typically requires massive amounts of manually labeled data – a slow, expensive process. Existing self-supervised approaches also struggle because they ignore the natural hierarchy of heartbeats and rhythms crucial for accurate diagnosis. We introduce MELP, a novel AI model that learns from unlabeled ECG-text pairs. MELP first deeply understands cardiology reports. It then aligns ECG signals with text descriptions at three clinically meaningful levels: fine-grained waveform features (token), individual heartbeats (beat), and overall rhythm patterns – mimicking how cardiologists interpret ECGs. MELP learns highly transferable ECG representations without expensive manual labels. It significantly outperforms prior methods across three public datasets, excelling at tasks like identifying new heart conditions without training ("zero-shot") and adapting efficiently to various ECG analysis tasks. This paves the way for faster, cheaper, and more adaptable AI tools for heart health monitoring.

\begin{abstract}
Electrocardiograms (ECGs) play a vital role in monitoring cardiac health and diagnosing heart diseases. However, traditional deep learning approaches for ECG analysis rely heavily on large-scale manual annotations, which are both time-consuming and resource-intensive to obtain. To overcome this limitation, self-supervised learning (SSL) has emerged as a promising alternative, enabling the extraction of robust ECG representations that can be efficiently transferred to various downstream tasks.
While previous studies have explored SSL for ECG pretraining and multi-modal ECG-language alignment, they often fail to capture the multi-scale nature of ECG signals. As a result, these methods struggle to learn generalized representations due to their inability to model the hierarchical structure of ECG data.
To address this gap, we introduce \modelname\, a novel \textbf{M}ulti-scale \textbf{E}CG-\textbf{L}anguage \textbf{P}retraining (\textbf{MELP}) model that fully leverages hierarchical supervision from ECG-text pairs. \modelname\ first pretrains a cardiology-specific language model to enhance its understanding of clinical text. It then applies three levels of cross-modal supervision—at the token, beat, and rhythm levels—to align ECG signals with textual reports, capturing structured information across different time scales.
We evaluate \modelname\ on three public ECG datasets across multiple tasks, including zero-shot ECG classification, linear probing, and transfer learning. Experimental results demonstrate that \modelname\ outperforms existing SSL methods, underscoring its effectiveness and adaptability across diverse clinical applications.
Our code is available at \url{https://github.com/HKU-MedAI/MELP}.

\end{abstract}
\section{Introduction}

Electrocardiograms (ECGs) are widely used for monitoring cardiac health and diagnosing cardiovascular diseases. The standard 12-lead ECG records electrical activity from different perspectives, capturing both temporal and spatial characteristics of the heart’s function.
Advances in deep learning have significantly improved the analysis of these signals, enhancing the analysis of their underlying patterns~\cite{yan2019fusing, ebrahimi2020review, siontis2021artificial}. 

\begin{figure}
    \centering
    \includegraphics[width=0.95\linewidth]{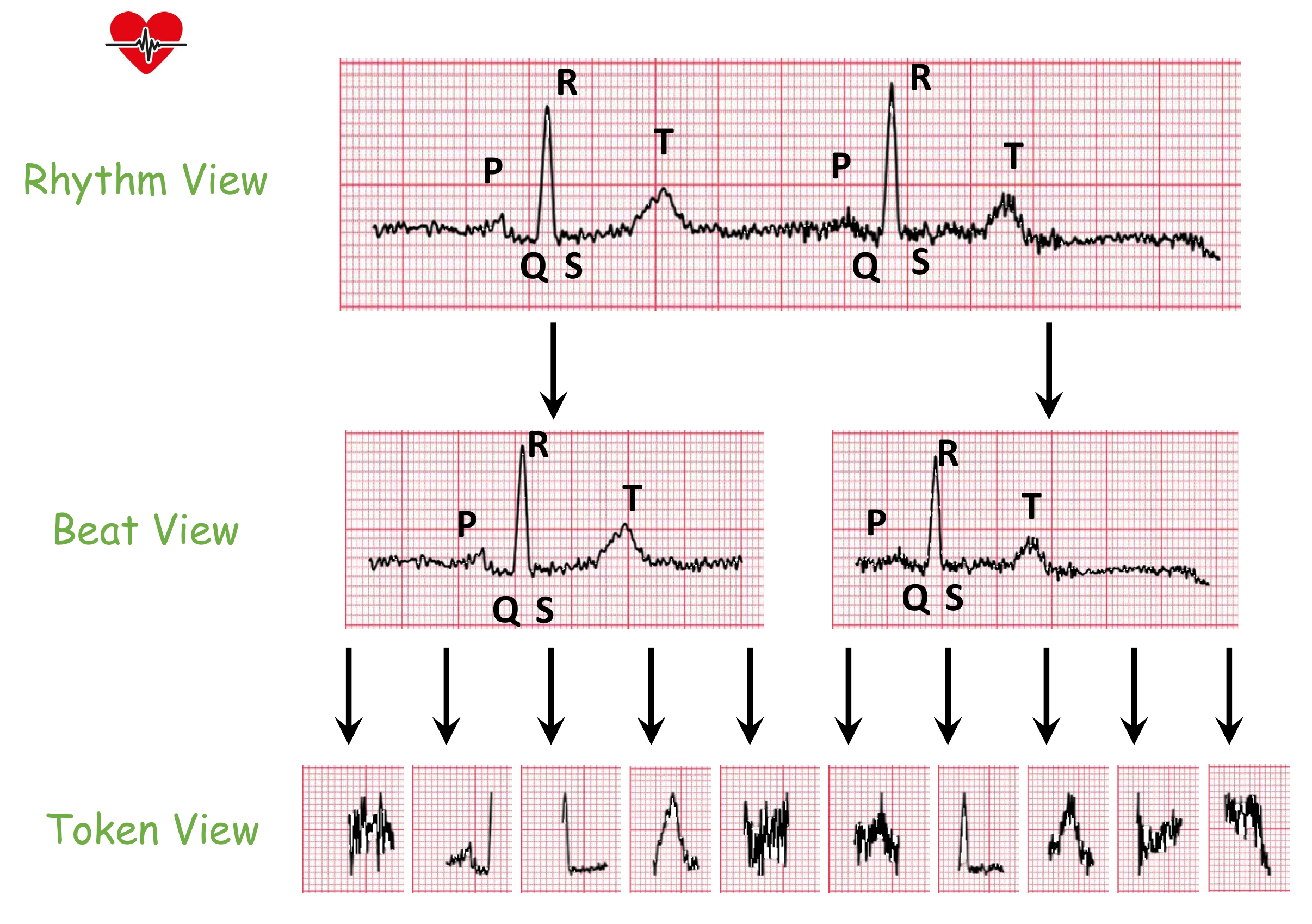}
    \vspace{-0.3cm}
    \caption{
    Illustration of the multi-scale view of ECG signals. \textbf{Rhythm Level}: Captures the full ECG recording, reflecting the heart’s global electrical activity over time. \textbf{Beats Level}: Segments each rhythm into discrete heartbeat tokens, isolating individual cardiac cycles for localized analysis. \textbf{Token Level}: Further decomposes each heartbeat into finer-grained temporal components, enabling granular feature extraction.  
    }
    \label{fig:teaser}
    \vspace{-0.6cm}
\end{figure}

Self-supervised learning (SSL) has emerged as a promising solution for ECG analysis, enabling meaningful representation learning without the need for labeled data. Existing SSL approaches for ECGs rely on either contrastive~\cite{kiyasseh2021clocs, oh2022lead, mckeen2024ecg} or generative methods~\cite{na2024guiding, hu2023spatiotemporal, jin2025reading}, but most focus solely on ECG signals without leveraging complementary clinical knowledge.
With the growing availability of clinical documentation, multi-modal learning—specifically ECG-language alignment—has gained attention. Recent studies~\cite{zhao2024ecg, yu2024ecg, pham2024cmelt} have explored methods for linking ECG signals with textual interpretations. Notably, ~\citet{liu2024merl} introduced MERL, a zero-shot ECG-language framework that uses contrastive learning to align ECG recordings with clinical reports.
Despite these advances, existing models primarily focus on global ECG-to-text alignment, overlooking the rich, multi-scale structure of ECG signals (Figure~\ref{fig:teaser}). 
For example, the rhythm-level view captures overall temporal rhythm of the ECG, the beat-level view analyzes the characteristics of each cardiac cycle, and token-level view interprets specific waveform segments.
All of these views are crucial for comprehensive ECG interpretation.
To address this limitation, a more detailed approach, incorporating token-, beat-, and rhythm-level representations, is essential for capturing comprehensive ECG features critical for precise diagnostics.
%
% Without this granularity, current methods struggle to fully leverage the semantic depth of clinical reports, limiting their effectiveness in real-world applications.

To address this gap, we propose \textbf{MELP}, a novel \textbf{M}ulti-scale \textbf{E}CG-\textbf{L}anguage \textbf{P}retraining model that fully exploits the multi-scale structure of ECGs at the token, beat, and rhythm levels, incorporating detailed cross-modal knowledge from clinical text. 
A key component of \modelname\ is a dedicated pretraining stage, where we first train a cardiology-specific language model before jointly training on paired ECG-text datasets. This step enhances the model’s ability to interpret medical terminology and align textual information with ECG signals.
To capture fine-grained modality interactions at the token level, we introduce an ECG captioning task, enabling the model to generate descriptive representations of short waveform segments. At the beat level, we extract heartbeat embeddings from token representations and sentence embeddings from word tokens, applying a contrastive learning objective to align beats with their corresponding clinical descriptions. Finally, at the rhythm level, we incorporate a global contrastive loss to learn robust representations of full ECG recordings.
This multi-scale approach bridges the gap between raw ECG signals and their clinical interpretations, allowing \modelname\ to learn highly transferable representations applicable across a range of tasks. We evaluate \modelname\ on three public ECG datasets, demonstrating its superiority over existing self-supervised and multi-modal models in zero-shot classification, linear probing, and cross-institutional transfer learning.

In summary, our contributions are threefold:
\begin{itemize}
    \item A novel Multi-scale ECG-Language Pretraining model (\modelname) that hierarchically integrates clinical text knowledge for improved ECG representation learning.
    \item A structured pretraining framework with explicit cross-modal supervision at three clinically meaningful levels: token, beat, and rhythm.
    \item Comprehensive evaluation on three public ECG datasets, achieving state-of-the-art performance in zero-shot classification, linear probing, and cross-domain transfer learning.
\end{itemize}

\section{Related Works}
\subsection{ECG Representation Learning}

Self-supervised learning (SSL) has demonstrated significant efficacy in leveraging unlabeled data across diverse domains, including natural language processing~\cite{devlin2018bert, dong2019unified, he2020deberta}, computer vision~\cite{wu2018unsupervised,grill2020bootstrap,caron2021emerging}, and time-series analysis~\cite{yue2022ts2vec, nie2022time, eldele2021time}. Recently, SSL has been extended to electrocardiogram (ECG) signal analysis, enabling robust representation learning for downstream pathology detection tasks~\cite{baevski2020wav2vec, gopal20213kg}. Existing methodologies in ECG SSL primarily fall into two categories: \textit{contrastive} and \textit{generative} approaches.

Contrastive methods ~\cite{sangha2024biometric}, currently the dominant paradigm, aim to maximize similarity between representations of augmented views of the same instance (positive pairs) while minimizing similarity with unrelated instances (negative pairs). Widely adopted frameworks such as SimCLR~\cite{chen2020simple} and MoCo~\cite{he2020momentum} have inspired ECG-specific adaptations. For example, Kiyasseh et al.~\cite{kiyasseh2021clocs} employ tailored signal augmentations (e.g., lead masking, noise) to generate positive ECG pairs for contrastive training.

Generative approaches, such as ST-MEM~\cite{na2024guiding} and HeartLang~\cite{jin2025reading}, learn representations by reconstructing masked portions of ECG signals. These methods ~\cite{yu2023ecg} typically occlude temporal segments (e.g., P-waves, T-waves) or entire beats and train models to recover the original waveform. HeartLang further introduces a tokenization strategy that segments ECG recordings by detecting QRS complexes, mapping these physiological events to discrete embeddings via a trainable codebook.

Recent efforts~\cite{oh2022lead, mckeen2024ecg, song2024foundation} combine contrastive and generative objectives to develop ECG foundation models. These frameworks are pre-trained on heterogeneous ECG datasets to learn generalizable representations. While promising, existing methods operate solely on uni-modal ECG data, overlooking the rich semantic correlations between ECG signals and clinical text reports, which is a limitation our work explicitly addresses.

\subsection{ECG-Language Pretraining}

% Multi-modal presentation learning, which aims to combine information from different modality by aligning representations from different modality. Outstanding prior arts ~\cite{radford2021learning} such as clip, which aligns image and text data by using contrastive methods. Furthermore, few works have introduced multi-modal representation learning into ECG data analysis. While one prior art~\cite{yu2024ecg} utilize Large Language Model (LLM)to rewrite EHR data and guide ECG representation learning, Merl ~\cite{liu2024merl} proposes a multi-modal model aligning ECG data and clinical text reports to acheive a good performance on zero-shot testing . A few more works proceed to combine text reports into analysis. C-melt ~\cite{pham2024cmelt} conducts generative approaches on both modalities, say Mask Language Modeling(MLM) and Mask ECG Modeling(MEM), and designs cross-modality contrastive learning, say ECG-Text alignment. ECG chat~\cite{zhao2024ecg} based on the structure of CoCa~\cite{yu2022coca}, one of SOTA works adpoting constrastive approaches, to achieve text report and ECG recording alignment. However, these models mainly focus on learning a single generalized ECG representation from the entire ECG recording and ignore local information of ECG recordings. Therefore, these models fail to extract local task specific features, which limits their performance on the diverse downstream tasks.

Multi-modal representation learning aims to integrate information from diverse modalities by aligning their respective representations. Notable prior works, such as CLIP~\cite{radford2021learning}, employ contrastive methods to align image and text data. However, relatively few studies ~\cite{li2024frozen,han2024foundation, tian2024foundation} have explored multi-modal representation learning in the context of ECG data analysis. For instance, ESI ~\cite{yu2024ecg} leverages a Large Language Model (LLM) to reinterpret electronic health record (EHR) data and guide ECG representation learning, while MERL~\cite{liu2024merl} proposes a framework that aligns ECG signals with clinical text reports to achieve strong zero-shot classification performance. Additional efforts, such as C-MELT~\cite{pham2024cmelt}, adopt generative approaches for both modalities—combining Masked Language Modeling (MLM) with Masked ECG Modeling (MEM)—and incorporate cross-modal contrastive learning for ECG-text alignment. Similarly, ECG-Chat~\cite{zhao2024ecg}, inspired by the CoCa architecture~\cite{yu2022coca} (a state-of-the-art contrastive framework), aligns ECG recordings with textual reports. 

Despite these advances, existing models predominantly focus on deriving a global  representation from the whole ECG recordings, overlooking fine-grained local patterns. Consequently, they do not effectively capture task-specific features at localized intervals, potentially limiting their adaptability to diverse downstream applications.

\section{Method}

\begin{figure*}[ht!]
    \centering
    \includegraphics[width=0.95\textwidth]{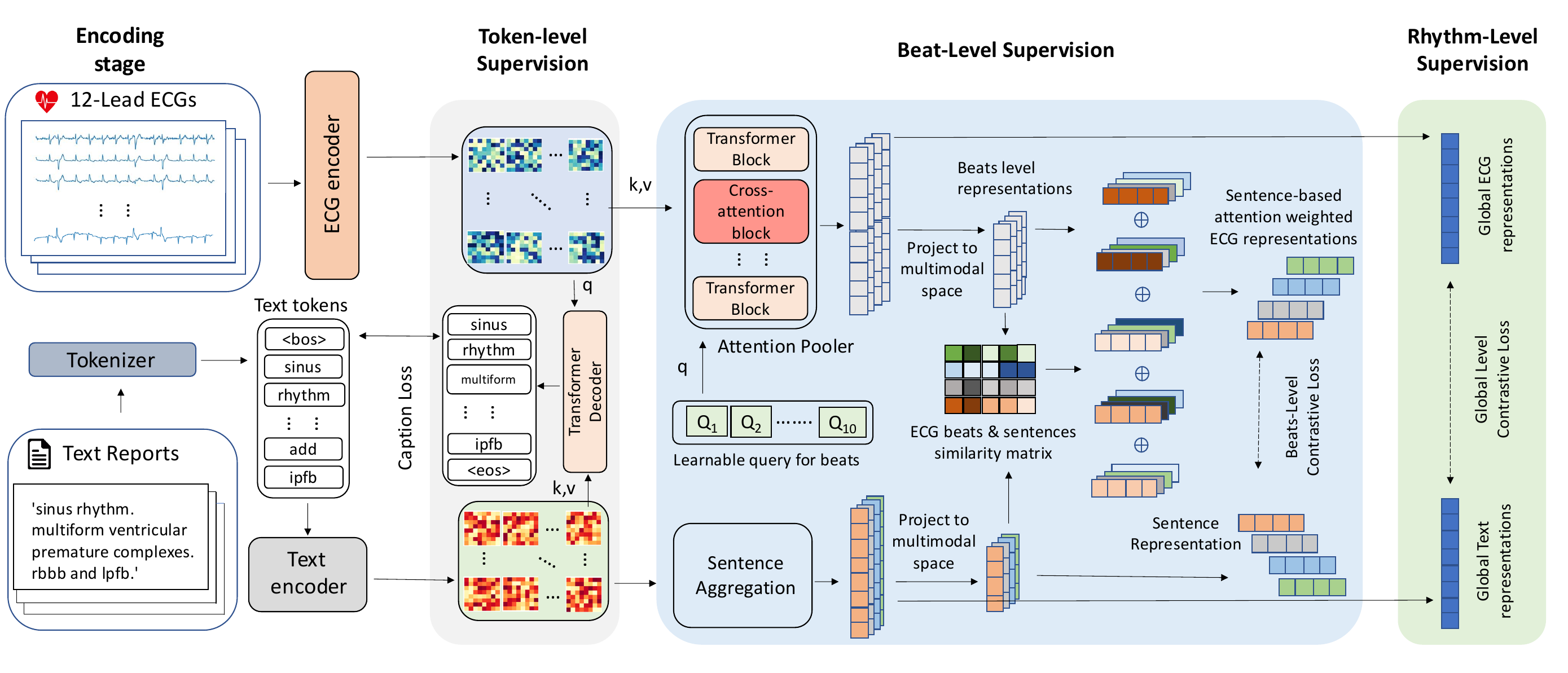}
    \vspace{-0.2cm}
    \caption{
    \textbf{Overview of \modelname}:
    \modelname\ incorporates three levels of supervision—token, beat, and rhythm—to guide ECG-language pretraining.
    At the token level, fine-grained ECG representations serve as queries for a transformer decoder, which reconstructs paired ECG reports using a captioning loss ($\mathcal{L}_{\mathrm{LM}}$). 
    At the beat level, token-level ECG features are aggregated into beat-level representations via an attention pooling layer, while text representations are grouped into sentence-level embeddings. A similarity matrix between ECG beats and text sentences is used to reweight these embeddings, optimizing a beat-level contrastive loss ($\mathcal{L}_{\mathrm{Local}}$). 
    At the rhythm level, beat-level ECG features and sentence-level text embeddings are further aggregated through average pooling to generate global representations, which are optimized using a global contrastive loss ($\mathcal{L}_{\mathrm{g}}$).
    }
    \label{fig:framework}
    \vspace{-0.4cm}
\end{figure*}

\subsection{Overview}

Figure~\ref{fig:framework} provides an overview of \modelname\, which learns generalized ECG representations through multi-scale cross-modal supervision in a self-supervised manner.
We begin by introducing cardiology language pretraining strategies (Section~\ref{sec:unimodal pretraining}) to equip the text encoders with domain-specific knowledge.
Next, we present our multimodal pretraining framework in Section~\ref{sec:multimodal pretraining}, which leverages fine-grained cross-modal supervision to align ECG and text modalities effectively.
Finally, we describe the evaluation protocol for transferring the pretrained framework to downstream tasks in Section~\ref{sec:evaluation}.

\subsection{Cardiology Language Pretraining}
\label{sec:unimodal pretraining}

As witnessed in the prior work~\cite{boecking2022making} in medical vision-language pretraining, the language plays a significant role in learning generalized visual representations. 
To maximize the language model’s utility for cardiology, we pretrain a text encoder using a cardiology-focused corpus, building on the MED-CPT query encoder~\cite{jin2023medcpt}.  
The corpus is curated from three sources: cardiology-related data from PubMed\footnote{https://www.ncbi.nlm.nih.gov/home/develop/api/}, Wikipedia\footnote{https://en.wikipedia.org/wiki/Wikipedia}, and the MIMIC-IV-ECG training set~\cite{gow2023mimic}, following the approach of HeartBERT~\cite{gwon2024medical}.  
We use the masked language modeling objective~\cite{devlin2018bert} to pretrain the model. 
% This enables the encoder to capture domain-specific knowledge effectively.  
%  
The details of pretraining the cardiology language model is presented in Appendix~\ref{sec:cardiology_language}.  

\subsection{Multimodal Pretraining}  
\label{sec:multimodal pretraining}  

\para{Motivation.}  
Cardiologists interpret ECG signals in a hierarchical manner, analyzing features at multiple scales—from individual waveform components (tokens) to heartbeats (beats) and overall rhythm. 
This hierarchical approach forms the basis for many clinical diagnostic criteria.
We observe that all three levels are essential for accurate ECG interpretation, as illustrated by the following examples (More examples are provided in Appendix \ref{sec:multi-level-design}):
\begin{itemize}
    \item Token-level: In diagnosing Atrial Fibrillation, clinicians look for absence of P waves and QRS duration usually $\textless120ms$. These features are localized within short waveform segments, corresponding to the token level~\cite{lb2011ccs}.
    \item Beat-level: For sinus rhythm, each QRS complex should be preceded by a normal P wave True. This criteria requires to analyze two waveform components together across the entire heart beat.~\cite{mattu2019electrocardiography}
    \item Rhythm-level: Diagnosing Left Anterior Fascicular Block (LAFB) involves identifying left axis deviation—e.g., negative deflections in leads II, III, and aVF, and positive deflections in leads I and aVL~\cite{mattu2019electrocardiography}. This assessment depends on recognizing patterns over the entire ECG lead.
\end{itemize}
\vspace{-0.4pt}
Inspired by this structured approach, our model, \modelname, integrates three levels of cross-modal supervision for ECG-language pretraining. 
Formally, we define the ECG encoder as $f(\cdot; \theta)$ and a text encoder $g(\cdot; \phi)$, where the ECG encoder is random initialized and the text encoder is initialized from the pretrained language model (Sec.~\ref{sec:unimodal pretraining}). 
Given an ECG sample $X$ and its paired ECG report $T$, we omit sample index ($i$)  for clarity.

\para{Token-view: Learning to Generate Captions.}
To leverage the benefits of multimodal generative pretraining, we adopt an encoder-decoder architecture for ECG-language learning. The text decoder is designed to generate ECG reports with fine-grained detail, predicting tokenized text autoregressively based on ECG token embeddings.
For each ECG sample $X$, our ECG encoder $f$ produces token-level ECG embedding $E \in \mathbb{R}^{L_{t} \times D}$, where $L_{t}$ is the number of ECG tokens and $D$ the feature dimension.
To summarize these embeddings, we use an attention pooler~\cite{yu2022coca} with 128 learnable query tokens, resulting in $\Tilde{E} \in \mathbb{R}^{128\times D}$.
The attention pooler consists of a single multi-head attention layer, where the encoder output serves as both keys and values.
Following a GPT-style autoregressive framework, the text decoder maximizes the conditional likelihood of the paired report through next-word prediction.
For a token sequence  $T = (<\mathrm{BOS}>, w_1, ..., w_N, <\mathrm{EOS}>)$, the text decoder optimizes:
\begin{equation}
    \mathcal{L}_{\mathrm{LM}}(\zeta) = -\sum_{i=1}^{N} \mathrm{log}p(w_i | w_{0:i-1}, \Tilde{E}) 
\end{equation}
where $\zeta$ denotes the learnable parameters of the text decoder.
The text decoder is randomly initialized and trained with teacher-forcing~\cite{williams1989learning} to enhance computational efficiency and learning speed.
An additional advantage of this generative pretraining approach is its flexibility in adapting to downstream tasks, such as ECG report generation and ECG Question Answering (ECG-QA).

\para{Discussion.}
While some ECG reports provide high-level rhythm summaries (e.g., ``sinus rhythm"), many also include detailed descriptions of waveform-level abnormalities. Examples of such observations are provided in Appendix~\ref{sec:multi-level-design}.
Besides, high-level findings, like ``sinus rhythm", still depend on low-level indicators such as P wave consistency and PR interval regularity. 
By providing full waveform features to the decoder, token-level pretraining allow the model to learn these relationships and generate reports with varying levels of granularity. It may encourage the model to analyze local features and implicitly learn these indicators.
\begin{table*}[t!]
    \centering
    \caption{Details on the number of samples in each split for each downstream dataset.}
    \resizebox{0.9\textwidth}{!}{
    \begin{tabular}{c c c c c c c c}
        \toprule
        \#. Samples & MIMIC-IV-ECG & PTBXL-Rhythm & PTBXL-Sub & PTBXL-Form & PTBXL-Super & CPSC2018 & CSN \\
        \midrule
        Train & 745,447 & 16,832 & 17,084 & 7,197 & 17,084 & 4,950 & 16,546 \\
        Validation & 15,171 & 2,100 & 2,146 & 901 & 2,146 & 551 & 1,860 \\
        Test & - & 2,098 & 2,158 & 880 & 2,158 & 1,376 & 4,620 \\
        \bottomrule
    \end{tabular}
    }
    \label{tab:dataset_statistics}
    \vspace{-0.4cm}
\end{table*}

\para{Beat view: Heart Beat-Sentence Alignment.}
ECG signals can be segmented into individual heart beats, allowing for more interpretable cardiological analysis~\cite{jin2025reading}. However, strictly relying on R-peak detection may disrupt inter-beat morphological information and temporal relationships. To mitigate this, we aggregate beat embeddings in the latent space using an attention pooler. 
Specifically, we introduce 10 learnable tokens to hierarchically summarize beats within a 10-second ECG segment.
While we use a default setting of 10 tokens in the main manuscript, an empirical analysis presented in Appendix~\ref{sec:heart_beats} shows that this is not necessarily the optimal choice.
We leave the exploration of more adaptive and clinically informed heartbeat alignment strategies for future work.

Clinical observations suggest that transient abnormal beats often correspond to specific sentences in ECG reports. To capture this fine-grained alignment, we propose a beat-sentence matching mechanism.
Let $B \in \mathbb{R}^{N_B\times D}$ represent beat embeddings obtained from the ECG encoder’s attention-pooled outputs, and let $S \in \mathbb{R}^{S\times D}$ denote sentence embeddings, where each sentence embedding is computed by averaging its word tokens.
A projection layer—either a linear transformation or a two-layer MLP—maps both embeddings into a shared latent space using functions $p_{\mathrm{E}}$ (for ECG) and $p_{\mathrm{T}}$(for Text).
For simplicity, we continue to denote the resulting embeddings as $B$ and $S$.
For each sentence embedding $S(l) \in \mathbb{R}^{D}$, we compute an attention-weighted beat embedding $\hat{B}(l)$ using this equation:
\begin{equation}
    \hat{B}(l) = \sum_{l=1}^{N_B} \alpha_{l} S(l)
\end{equation}
where the attention weight $\alpha_{l}$ is defined as:
\begin{equation}
    \alpha(l) = \frac{\mathrm{exp}(\langle S(l), B(l) \rangle / \tau_1)}{\sum_{j=1}^{N_B} \mathrm{exp}(\langle S(l), B(j)\rangle / \tau_1)}
\end{equation}
Here $\langle \cdot \rangle$ denotes cosine similarity, $\tau_1 = 0.25$ is a temperature hyperparameter.
The similarity between an ECG-text pair is then computed by aggregating similarities between attention-weighted beat embeddings and the corresponding text embedding:
\begin{equation}
    Z(X, T) = \mathrm{log}\Big(\sum_{l=1}^{S}\mathrm{exp}(\langle \hat{B}(l), S(l) \rangle) / \tau_2 \Big)^{\tau_2}
\end{equation}
where $\tau_2 = 0.1$ is another temperature hyperparameter.

To optimize alignment, we define the local contrastive loss for a minibatch of size B as:
\begin{equation}
    \mathcal{L}_{\mathrm{Local}}^{e\rightarrow t} = \frac{1}{B}\sum_{i=1}^{B} -\mathrm{log}\Big( 
    \frac{\mathrm{exp}(Z(X_i, T_i)) / \tau_2}{\sum_{k=1}^{B} \mathrm{exp}(Z(X_i, T_k)) / \tau_2} \Big)
\end{equation}
where $i$ is the sample index.

The total local loss is then given by: $\mathcal{L}_{\mathrm{Local}} = \frac{1}{2} (\mathcal{L}_{\mathrm{Local}}^{e\rightarrow t} + \mathcal{L}_{\mathrm{Local}}^{t\rightarrow e})$.
This loss function ensures robust heartbeat-sentence alignment, preserving both local and global ECG-text relationships.

\para{Rhythm view: ECG-Report Alignment.}
Inspired by the success of CLIP~\cite{radford2021learning}, contrastive loss has proven effective in learning transferable multimodal representations. Following this principle, we introduce an instance-level contrastive loss to establish high-level cross-modal supervision between ECG signals and text reports.

To obtain global representations, we compute the ECG embedding by averaging all beat embeddings, while the text embedding is represented by the [CLS] token. For the $i$-th ECG-text pair, we denote these global embeddings as  $X_i^g$ (ECG) and $T_i^g$ (text). 
Their similarity is computed using cosine similarity as $\langle X_i^g, T_i^g \rangle$.
The global alignment loss $\mathcal{L}_\mathrm{g} = \frac{1}{2}(\mathcal{L}_{\mathrm{g}}^{e \rightarrow t} + \mathcal{L}_{\mathrm{g}}^{t \rightarrow e})$.
The $\mathcal{L}_{\mathrm{g}}^{e \rightarrow t}$ is defined following the InfoNCE formulation~\cite{oord2018representation}:
\begin{equation}
\mathcal{L}_{\mathrm{g}}^{e \rightarrow t} = -\frac{1}{B}\log \frac{\exp(\langle X_i^g, T_i^g \rangle / \tau)}{\sum_{j=1}^B \exp(\langle X_i^g, T_j^g \rangle / \tau)}
\end{equation}
where $\tau$ is a learnable temperature hyperparameter, and B denotes the batch size.

\para{Overall Pretraining.}
Our framework is trained by jointly optimizing three loss functions:
\begin{equation}
    \mathcal{L} = \mathcal{L}_{\mathrm{g}} + \lambda_1 * \mathcal{L}_{\mathrm{LM}} + \lambda_2
 * \mathcal{L}_{\mathrm{Local}}
\end{equation}
where $\lambda_1$ and $\lambda_2$ control the contributions of the language modeling and local alignment losses, respectively.
Based on empirical results, we set $\lambda_1 = 2$ and $\lambda_2 = 0.2$ for pretraining.
The experimental results can be found in Sec.~\ref{sec:ablation_hyperparameters}.

\subsection{Transferring into Downstream Tasks}
\label{sec:evaluation}
For zero-shot evaluation, we directly use the global embeddings $X_i^g$ (ECG) and $T_i^g$ (Text) for retrieval-based tasks.

For fine-tuning, we first extract beat-level embeddings before the projection layer $p_E$.
These embeddings are then aggregated via average pooling to obtain a global feature vector. A linear classification layer is added on top to generate per-class predictions.

In the linear probing setup, we freeze the entire network and train only the linear layer. 
% Further details on category alignment and domain adaptation protocols can be found in Appendix~\ref{sec:pretraining_details}.
\section{Experiments}

\colorlet{shadegray}{gray!20}
\definecolor{darkgreen}{rgb}{0,0.5,0}

\begin{table*}[t!]
    \centering
    \caption{
    Linear probing performance (AUC [\%]) of \modelname\ and baseline models across multiple datasets.
    Results are reported for different training data proportions (1\%, 10\%, and 100\%).
    The best and \underline{second-best} results are highlighted in bold and \underline{underlined}, respectively.
    }
    \label{tab:linear_probe}
    \resizebox{0.95\textwidth}{!}{
        \begin{tabular}{cccc|ccc|ccc|ccc|ccc|ccc}
        \toprule
        Methods  & \multicolumn{3}{c}{PTBXL-Rhythm} &  \multicolumn{3}{c}{PTBXL-Sub} & \multicolumn{3}{c}{PTBXL-Form} 
         & \multicolumn{3}{c}{PTBXL-Super} & \multicolumn{3}{c}{CPSC2018} & \multicolumn{3}{c}{CSN} \\
        \midrule
        Training ratio& 1\% & 10\% & 100\% & 1\% & 10\% & 100\% & 1\% & 10\% & 100\% 
        & 1\% & 10\% & 100\% & 1\% & 10\% & 100\% & 1\% & 10\% & 100\%\\
        \midrule
        SimCLR~\cite{chen2020simple}      
         & 51.41 & 69.44 & 77.73  & 60.84 & 68.27 & 73.39 & 54.98 & 56.97 & 62.52 & 63.41 & 69.77 & 73.53 & 59.78 & 68.52 & 76.54 & 59.02 & 67.26 & 73.20\\
        BYOL~\cite{grill2020bootstrap}         
         & 41.99 & 74.40 & 77.17  & 57.16 & 67.44 & 71.64 & 48.73 & 61.63 & 70.82 & 71.70 & 73.83 & 76.45 & 60.88 & 74.42 & 78.75 & 54.20 & 71.92 & 74.69\\
        BarlowTwins~\cite{zbontar2021barlow}  
         & 50.12 & 73.54 & 77.62  & 62.57 & 70.84 & 74.34 & 52.12 & 60.39 & 66.14& 72.87 & 75.96 & 78.41 & 55.12 & 72.75 & 78.39 & 60.72 & 71.64 & 77.43\\
        MoCo-v3~\cite{chen2021empirical}      
         & 51.38 & 71.66 & 74.33  & 55.88 & 69.21 & 76.69 & 50.32 & 63.71 & 71.31& 73.19 & 76.65 & 78.26 & 62.13 & 76.74 & 75.29 & 54.61 & 74.26 & 77.68\\
        SimSiam~\cite{chen2021exploring}      
         & 49.30 & 69.47 & 75.92  & 62.52 & 69.31 & 76.38 & 55.16 & 62.91 & 71.31 & 73.15 & 72.70 & 75.63 & 58.35 & 72.89 & 75.31 & 58.25 & 68.61 & 77.41\\
        TS-TCC~\cite{eldele2021time}      
         & 43.34 & 69.48 & 78.23  & 53.54 & 66.98 & 77.87 & 48.04 & 61.79 & 71.18& 70.73 & 75.88 & 78.91 & 57.07 & 73.62 & 78.72 & 55.26 & 68.48 & 76.79\\
        CLOCS~\cite{kiyasseh2021clocs}       
         & 47.19 & 71.88 & 76.31  & 57.94 & 72.55 & 76.24 & 51.97 & 57.79 & 72.65& 68.94 & 73.36 & 76.31 & 59.59 & 77.78 & 77.49 & 54.38 & 71.93 & 76.13\\
        Wav2Vec 2.0 + CMSC + RLM ~\cite{oh2022lead}
        & 76.24 & 86.34 & 92.05 & 69.10 & 80.71 & 85.01 & 52.72 & 67.81 & 80.72 & 81.15 & 84.88 & 85.53 & 75.70 & 88.16 & 92.61 & 65.65 & 78.82 & 87.87\\
        ASTCL~\cite{10177892}        
         & 52.38 & 71.98 & 76.05  & 61.86 & 68.77 & 76.51 & 44.14 & 60.93 & 66.99 & 72.51 & 77.31 & 81.02 & 57.90 & 77.01 & 79.51 & 56.40 & 70.87 & 75.79\\
        CRT~\cite{zhang2023self}         
         & 47.44 & 73.52 & 74.41  & 61.98 & 70.82 & 78.67 & 46.41 & 59.49 & 68.73& 69.68 & 78.24 & 77.24 & 58.01 & 76.43 & 82.03 & 56.21 & 73.70 & 78.80\\
         ECGFM~\cite{mckeen2024ecg} 
         & \underline{81.45} & \underline{91.59} & \underline{92.70}  & \underline{73.24} & \underline{81.91} & 86.07 & \underline{60.95} & \underline{74.99} & \textbf{85.54} & 78.67 & 84.80 & 86.47 & \underline{82.18} & \underline{89.52} & \underline{93.26} & \underline{71.51} & \underline{83.17} & \underline{88.89} \\
        ST-MEM~\cite{na2024guiding}       
         & 51.12 & 65.44 & 74.85  & 54.12 & 57.86 & 63.59 & 55.71 & 59.99 & 66.07& 61.12 & 66.87 & 71.36 & 56.69 & 63.32 & 70.39 & 59.77 & 66.87 & 71.36\\
        HeartLang~\cite{jin2025reading} 
        & 62.08 & 76.22 & 90.34 &64.68 &79.34 & \textbf{88.91} & 58.70 & 63.99 & 80.23 & 78.94 &85.59 &87.52 &60.44 &66.26 &77.87 & 57.94 &68.93 &82.49\\
        MERL~\cite{liu2024merl}         
         & 53.33 & 82.88 & 88.34 & 64.90 & 80.56 & 84.72 & 58.26 & 72.43 & 79.65 & \underline{82.39} & \underline{86.27} & \textbf{88.67} & 70.33 & 85.32 & 90.57 & 66.60 & 82.74 & 87.95 \\
        % C-MELT~\cite{pham2024cmelt}  &86.61 &92.83 &96.71  &77.74 &82.92 &85.15 &\textbf{70.10} &\textbf{78.91} &\textbf{83.98}&83.15 &\textbf{88.36} &\textbf{90.11} &85.46 &91.35& \textbf{94.92} & \textbf{80.04} &\textbf{87.36} &\textbf{90.71}\\
        \midrule
        \rowcolor{shadegray}
        \modelname\ (Ours)   & \textbf{88.83} & \textbf{94.65} & \textbf{96.91}  & \textbf{79.22} & \textbf{84.40} & \underline{87.46} & \textbf{63.41} & \textbf{76.71} & \underline{83.30} & \textbf{85.82} & \textbf{87.61} & \underline{87.87}  &\textbf{88.54} & \textbf{91.75} & \textbf{94.32} & \textbf{78.25} & \textbf{84.83} & \textbf{90.17} \\
        \bottomrule 
        \end{tabular} 
    }
    \vspace{-0.2cm}
\end{table*}

\begin{table*}[ht!]
    \centering
    \caption{
    Zero-shot classification performance (AUC [\%]) of \modelname\ and baseline models across multiple datasets.
    }
    \label{tab:zero_shot}
    \resizebox{0.95\textwidth}{!}{
    \begin{tabular}{cccccccc}
        \toprule
        Methods & CSN   & PTBXL-Rhythm &  PTBXL-Form & PTBXL-Sub & PTBXL-Super&  CPSC2018  & Average \\
        \midrule
        MERL~\cite{liu2024merl}    & 74.4        & 78.5      & 65.9       & 75.7         & 74.2     & 82.8 & 75.3    \\
        % C-MELT~\cite{pham2024cmelt}  & 76.3        & 88.6      & 66.1       & 75.9         & 76.2     & 80.1 & 77.1  \\
        \midrule
        \rowcolor{shadegray}
        \modelname\ (Ours) & \textbf{77.6} & \textbf{85.4} & \textbf{69.1} & \textbf{81.2} & \textbf{76.2} & \textbf{84.2} & \textbf{79.0} \\
        \highlightgain{Gains} & \highlightgain{+3.2} & \highlightgain{+6.9} & \highlightgain{+3.2} & \highlightgain{+5.5} & \highlightgain{+2.0} & \highlightgain{+1.4} & \highlightgain{+3.7} \\
        \bottomrule
    \end{tabular}
    }
    \vspace{-0.4cm}
\end{table*}
%
% \begin{table*}
%     \centering
%     \caption{ECG interpretation comparison.}

%     \resizebox{\textwidth}{!}{
%     \begin{tabular}{c c c c c c}
%         \toprule
%         Cardio Resident & Emergency Resident & Medical Student & DNN~\cite{ribeiro2020DNN} & C-MELT~\cite{pham2024cmelt} & \modelname \\
%         \midrule
%         92.07 & 90.50 & 93.61 & 96.59 & 96.79 & \\
%         \bottomrule
%     \end{tabular}
%     }
% \end{table*}

\subsection{Experimental Setup}
\subsubsection{Pretraining Task}
\para{Pretraining Dataset.} 
For the pretraining stage, we utilize the \textit{MIMIC-IV-ECG v1.0 database}~\cite{gow2023mimic}, comprising 800,035 ECG recordings from 161,352 unique patients.  
Each recording consists of a 10-second waveform sampled at 500 Hz.  
The database provides multimodal alignment through clinical text reports, with up to 18 textual reports paired with each ECG recording.  
We adopt preprocessing protocols adapted from~\cite{liu2024merl}, including text normalization (e.g., lowercase conversion, punctuation removal, and special character elimination).  
To ensure data quality, we exclude: (1) ECG recordings containing empty or NaN (Not-a-Number) values, (2) text reports with fewer than four tokens, and (3) ECG samples lacking paired textual annotations.  
This rigorous curation process results in a final dataset of 760,618 high-quality ECG-text pairs. 

\para{Implementation Details.}
The ECG encoder is based on the Wav2Vec 2.0 architecture~\cite{baevski2020wav2vec}, which integrates a multi-layer convolutional neural network (CNN) feature extractor with a transformer encoder. The ablation results justifying our selection of the Wav2Vec 2.0 architecture as the ECG encoder are included in Section \ref{sec:ablation on ecg encoder}.
We use the AdamW optimizer with an initial learning rate of 2e-4, a weight decay of 0.2, and a cosine annealing learning rate scheduler. 
\modelname\ is pretrained for 100 epochs with a per-device batch size of 64. 
Training is stopped early if the zero-shot prediction performance on the validation sets does not improve for five consecutive epochs. 
Please refer to our code for more details.
All experiments are conducted on four NVIDIA GTX 3090 GPUs.

\begin{table*}[t!]
    \centering
    \caption{
    Performance under data distribution shift.     
    ``Source Domain" refers to the dataset used for linear probing with the frozen pre-trained ECG encoder, while ``Target Domain" represents the corresponding test set.
    The \textbf{Best} and \underline{Second-best} results are shown in \textbf{Bold} and \underline{underlined}.
    }
    \label{tab:transfer}
    \resizebox{0.95\textwidth}{!}{
        \begin{tabular}{c c c c c c c c c c}
        \toprule
        Source Domain & \multirow{2}{*}{Zero-shot} & \multirow{2}{*}{Training Ratio} 
        & \multicolumn{2}{c}{PTBXL-Super}
        & \multicolumn{2}{c}{CPSC2018}
        & \multicolumn{2}{c}{CSN} 
        & \multirow{2}{*}{Average} \\
        \cline{4-5}
        \cline{6-7}
        \cline{8-9}
        Target Domain & ~ & ~ 
        & CPSC2018 & CSN & PTBXL-Super & CSN & PTBXL-Super & CPSC2018 \\
        \midrule
        SimCLR~\cite{chen2020simple}      
        &\XSolidBrush & 100\% & 69.62 & 73.05 & 56.65 & 66.36 & 59.74 & 62.11 & 65.22 \\
        BYOL~\cite{grill2020bootstrap}      
        &\XSolidBrush & 100\% & 70.27 & 74.01 & 57.32 & 67.56 & 60.39 & 63.24 & 65.63 \\
        BarlowTwins~\cite{zbontar2021barlow} 
        &\XSolidBrush & 100\% & 68.98 & 72.85 & 55.97 & 65.89 & 58.76 & 61.35 & 64.13 \\
        MoCo-v3~\cite{chen2021empirical}     
        &\XSolidBrush & 100\% & 69.41 & 73.29 & 56.54 & 66.12 & 59.82 & 62.07 & 64.21 \\
        SimSiam~\cite{chen2021exploring}     
        &\XSolidBrush & 100\% & 70.06 & 73.92 & 57.21 & 67.48 & 60.23 & 63.09 & 65.33 \\
        TS-TCC~\cite{eldele2021time}      
        &\XSolidBrush & 100\% & 71.32 & 75.16 & 58.47 & 68.34 & 61.55 & 64.48 & 66.55 \\
        CLOCS~\cite{kiyasseh2021clocs}       
        &\XSolidBrush & 100\% & 68.79 & 72.64 & 55.86 & 65.73 & 58.69 & 61.27 & 63.83 \\
        ASTCL~\cite{10177892}       
        &\XSolidBrush & 100\% & 69.23 & 73.18 & 56.61 & 66.27 & 59.74 & 62.12 & 64.19 \\
        CRT~\cite{zhang2023self}         
        &\XSolidBrush & 100\% & 70.15 & 74.08 & 57.39 & 67.62 & 60.48 & 63.33 & 65.51 \\
        ST-MEM~\cite{na2024guiding}      
        &\XSolidBrush & 100\% & 76.12 & \textbf{84.5} & 62.27 & 75.19 & 73.05 & 64.66 & 72.63 \\
        MERL~\cite{liu2024merl}        
        &\Checkmark & 0\%   & \textbf{88.21} & \underline{78.01} & \underline{76.77} & \underline{76.56} & \underline{74.15} & \textbf{82.86} & \underline{79.42} \\
        \midrule
        \rowcolor{shadegray}
        \modelname\ (Ours) & \Checkmark &  0\%  & \underline{87.75} & 74.11 & \textbf{77.89} & \textbf{80.32} & \textbf{74.67} & \underline{82.72} & \textbf{79.58} \\
        \bottomrule
        \end{tabular}
    }
    \vspace{-0.4cm}
\end{table*}

\subsubsection{Downstream Tasks}

\textbf{Downstream Datasets.} 
We evaluate our pre-trained \modelname\ across three publicly available benchmarks:  
PTB-XL~\cite{wagner2020ptb}, CSN~\cite{zheng2022large},  
CPSC2018~\cite{liu2018open}.  
A summary of dataset statistics is presented in Table~\ref{tab:dataset_statistics}, with additional details in Appendix~\ref{sec:dataset_details}. 
Key characteristics of these datasets are summarized below:  

\textit{PTB-XL} comprises 21,837 12-lead ECG recordings from 18,885 patients, each sampled at 500 Hz with a 10-second duration and annotated with cardiac diagnostic labels. Following the methodology of MERL~\cite{liu2024merl}, we stratify the dataset into four subgroups (super, sub, form, and rhythm) for granular evaluation. Training, validation, and test splits adhere to the protocol established by~\cite{wagner2020ptb}.  

\textit{CPSC2018} This resource contains 6,877 standard 12-lead ECG recordings sampled at 500 Hz, annotated with 9 categorical labels. We replicate the experimental configuration of MERL~\cite{liu2024merl} for consistency.  

\textit{CSN} contains 23,026 samples recorded at 500 Hz over 10-second intervals, this dataset includes 38 distinct diagnostic labels. For downstream evaluation, we adopt the train-validation-test partitioning scheme proposed by MERL~\cite{liu2024merl}.

\para{Implementation Details.}
For zero-shot evaluation, we use prompts driven by knowledge of GPT-4, following the approach in~\cite{liu2024merl}.
Linear probing tasks adhere strictly to the predefined train-validation-test splits from~\cite{liu2024merl}.
We conducted linear probing using 1\%, 10\% and 100\% of the training data for each task following~\cite{liu2024merl}.
All downstream tasks are evaluated using AUROC (Area Under the Curve). 
We use a batch size of 128 and train for 50 epochs, with early stopping similarly triggered based on the validation AUC.
Further implementation details are provided in Appendix~\ref{sec:pretraining_details}.
\begin{table*}[t!]
    \centering
    \caption{
    Ablation results of loss functions on 6 linear probing tasks.
    The first row indicates training with only the instance-level contrastive loss $\mathcal{L}_{\mathrm{g}}$.
    The \textbf{Best} and \underline{Second-best} results are shown in \textbf{Bold} and \underline{underlined}.
    }
    \label{tab:ablation_of_loss}
    \resizebox{0.95\textwidth}{!}{
    \begin{tabular}{c c c | c c c c c c c c c c c c c c c c c c | c}
        \toprule
        \multirow{2}{*}{$\mathcal{L}_{\mathrm{g}}$} & \multirow{2}{*}{$\mathcal{L}_{\mathrm{LM}}$} & \multirow{2}{*}{$\mathcal{L}_{\mathrm{Local}}$} &             
        \multicolumn{3}{c}{PTBXL-Rhythm}  & \multicolumn{3}{c}{PTBXL-Form}  & \multicolumn{3}{c}{PTBXL-Sub} &  
            \multicolumn{3}{c}{PTBXL-Super} & \multicolumn{3}{c}{CPSC2018} & \multicolumn{3}{c}{CSN} & \multirow{2}{*}{Average}  \\
        & & & 1\% & 10\% & 100\%& 1\% & 10\% & 100\%& 1\% & 10\% & 100\% & 1\% & 10\% & 100\% & 1\% & 10\% & 100\% & 1\% & 10\% & 100\% \\
        \midrule
        \checkmark & & &  83.78 & 88.44 & 94.98 & \underline{57.93} & 72.14 & 82.07 & 77.32 & 81.97 & 84.36 & 84.55 & 87.24 & 87.52 & 78.52 & 87.07 & 92.57 & \underline{75.94} & \underline{82.04} & 86.66 & 82.51 \\
        & \checkmark & & 77.64 & 79.44 & 85.21 & 52.95 & 63.80 & 76.91 & 71.41 & 76.67 & 82.97 & 78.73 & 82.80 & 85.18 & 64.19 & 73.05 & 85.26 & 69.81 & 79.37 & 84.41 & 76.10\\
        & & \checkmark  & 81.04 & \underline{89.88} & \underline{96.67} & 49.81 & 67.82 & 81.41 & 66.14 & 81.38 & 84.76 & 79.94 & \underline{87.49} & \underline{87.73} & 64.18 & 84.08 & \underline{93.17} & 55.89 & 79.77 & \underline{88.79} & 78.89 \\
        \checkmark & \checkmark & & 83.25 & 89.87 & 94.86 & 56.58 & \underline{72.71} & 81.99 & 78.61 & 82.14 & \underline{85.84} & 84.62 & 87.18 & 87.56 & \underline{83.74} & \underline{88.40} & 92.77 & 74.86 & 80.48 & 87.11 & \underline{82.92} \\
        \checkmark & & \checkmark &  \underline{84.36} & 88.44 & 95.29 & 57.22 & 72.07 & \underline{82.96} & \textbf{81.20} & \underline{82.89} & 85.42 & \underline{84.80} & 87.25 & 87.57 & 76.97 & 86.31 & 92.26 & 73.77 & 81.43 & 81.50 &82.32\\
        \checkmark & \checkmark & \checkmark & \textbf{88.83} & \textbf{94.65} & \textbf{96.91} & \textbf{63.41} & \textbf{76.71 }& \textbf{83.30} & \underline{79.22} & \textbf{84.40} & \textbf{87.46} & \textbf{85.82} & \textbf{87.61} & \textbf{87.87}  & \textbf{88.54} & \textbf{91.75} & \textbf{94.32} & \textbf{78.25} & \textbf{84.83} & \textbf{90.17}  & \textbf{85.78} \\
        \bottomrule
    \end{tabular}
    }
    \vspace{-0.4cm}
\end{table*}
\subsection{Quantitative Results}
\subsubsection{Evaluation on Linear Probing for ECG Classification}
Table~\ref{tab:linear_probe} presents the linear probing results comparing \modelname\ with baseline methods. 
We evaluate our \modelname\ against both unimodal self-supervised approaches, including TS-TCC~\cite{eldele2021time}, CLOCS~\cite{kiyasseh2021clocs}, ASTCL~\cite{10177892}, CRT~\cite{zhang2023self}, ST-MEM~\cite{na2024guiding}, and HeartLang~\cite{jin2025reading}, as well as multimodal self-supervised method such as MERL~\cite{liu2024merl}.
\modelname\ consistently improves classification performance across six tasks, achieving the highest accuracy in 16 out of 18 evaluation settings and ranking second in the remaining two. Its advantage is particularly evident when using only 1\% of training data, where it surpasses the second-best method in AUROC by margins of $+7.38\%$, $+5.98\%$, $+2.46\%$, $+3.43\%$, $+6.36\%$ and $+6.74\%$ respectively. 
These results underscore \modelname's effectiveness in low-data scenarios, highlighting its potential for real-world applications where labeled medical data is scarce.
With full training data (100\%), \modelname\ continues to rank first or second for all datasets , improving AUROC by $+4.21\%$, $+1.06\%$, and $+1.28\%$ on PTBXL-Rhythm, CPSC2018, and CSN, respectively. This demonstrates \modelname’s ability to achieve superior performance even with abundant labeled data, setting a strong upper bound for ECG classification. 
\begin{table*}[t!]
    \centering
    \caption{Ablation results of loss functions on 6 zero-shot classification tasks.}
    \label{tab:ablation_zero_shot}
    \resizebox{0.75\textwidth}{!}{
    \begin{tabular}{c c | c c c c c c | c}
        \toprule
        $\mathcal{L}_{\mathrm{LM}}$ & $\mathcal{L}_{\mathrm{Local}}$ &   PTBXL-Rhythm  & PTBXL-Form  & PTBXL-Sub &  
           PTBXL-Super & CPSC2018 & CSN & Average  \\
        \midrule
        & & 79.6 & \underline{68.7} & 77.1 & \textbf{77.2} & \underline{83.9} & 75.0 & \underline{76.9} \\
        \checkmark & & \underline{84.1} & 67.4 & 74.0 & 74.1 & 81.9 & \underline{76.8} & 76.4 \\
        & \checkmark & 82.8 & 64.4 & \underline{79.0} & \underline{75.9} & 81.1 & \underline{76.9}  & 76.7 \\
        \checkmark & \checkmark & \textbf{85.4} & \textbf{69.1} & \textbf{81.2} & 76.2 & \textbf{84.2} & \textbf{77.6} & \textbf{79.0} \\
        \bottomrule
    \end{tabular}
    }
    \vspace{-0.3cm}
\end{table*}
\begin{table*}[t!]
    \centering
    \caption{
    Ablation results of ECG encoders.
    We have used RLM as the augmentation technique for ECG by default.
    CMSC can't easily integrate into our model since it needs to split the ECG into two parts and performs contrastive learning.
    }
    \label{tab:ablation_of_ecg_encoder}
    \resizebox{0.95\textwidth}{!}{
        \begin{tabular}{ c | c c c c c c c c c c c c c c c c c c | c}
            \toprule
            \multirow{2}{*}{ECG encoder} 
             & \multicolumn{3}{c}{PTBXL-Rhythm}  & \multicolumn{3}{c}{PTBXL-Form} & \multicolumn{3}{c}{PTBXL-Sub} &  
            \multicolumn{3}{c}{PTBXL-Super} & \multicolumn{3}{c}{CPSC2018} & \multicolumn{3}{c}{CSN} & \multirow{2}{*}{Average}  \\
             & 1\% & 10\% & 100\% & 1\% & 10\% & 100\% & 1\% & 10\% & 100\% & 1\% & 10\% & 100\% & 1\% & 10\% & 100\% & 1\% & 10\% & 100\% \\
            \midrule 
            ResNet-18 & 85.10 & 90.11 & 94.31 & 62.82 & 73.59 & 79.23 & 75.59 & 81.72 & 85.70 & \textbf{85.84} & 86.99 & 87.24  & 83.31 & 89.75 & 93.35 &68.79 & 82.12 & 89.71& 83.07\\
             Wav2Vec 2.0  & \textbf{88.83} & \textbf{94.65} & \textbf{96.91} & \textbf{63.41} & \textbf{76.71} & \textbf{83.30} & \textbf{79.22} & \textbf{84.40} & \textbf{87.46} & 85.82 & \textbf{87.61} & \textbf{87.87}  & \textbf{88.54} & \textbf{91.75} & \textbf{94.32} &\textbf{78.25} & \textbf{84.83} & \textbf{90.17}& \textbf{85.78} \\
            Wav2Vec 2.0 + CMSC & 83.15 & 88.25 & 94.82 & 62.07 & 75.55 & 82.57 & 77.21 & 82.29 & 84.85  & 85.14 & 87.52 & 87.64 & 80.69 & 88.40 & 92.91 & 71.89 & 81.00 & 87.42 & 82.97\\
            \bottomrule
        \end{tabular}
    }
    \vspace{-0.3cm}
\end{table*}
\begin{table*}[t!]
    \centering
    \caption{
    Ablation results of pretrained domain-specific language model on 6 linear probing tasks.}
    \label{tab:ablation_of_text}
    \resizebox{0.95\textwidth}{!}{
        \begin{tabular}{ c | c c c c c c c c c c c c c c c c c c | c}
            \toprule
            \multirow{2}{*}{Text} 
             & \multicolumn{3}{c}{PTBXL-Rhythm}  & \multicolumn{3}{c}{PTBXL-Form}  & \multicolumn{3}{c}{PTBXL-Sub} &  
            \multicolumn{3}{c}{PTBXL-Super} & \multicolumn{3}{c}{CPSC2018} & \multicolumn{3}{c}{CSN} & \multirow{2}{*}{Average}  \\
             & 1\% & 10\% & 100\% & 1\% & 10\% & 100\% & 1\% & 10\% & 100\% & 1\% & 10\% & 100\% & 1\% & 10\% & 100\% & 1\% & 10\% & 100\% \\
            \midrule 
              &88.61 & \textbf{94.95} & 96.72 & \textbf{65.86} & 75.34& 82.94 & \textbf{80.36} & 84.17 & 86.43 &\textbf{86.08} & 87.46 & 87.77 & 86.93 & 91.33 & \textbf{93.48} & 75.82 & 83.95 & 89.51& 85.43 \\
             \checkmark   & \textbf{88.83} & 94.65 & \textbf{96.91} & 63.41 & \textbf{76.71} & \textbf{83.30} & 79.22 & \textbf{84.40} & \textbf{87.46} & 85.82 & \textbf{87.61} & \textbf{87.87}  & \textbf{88.54} & \textbf{91.75} & 94.32 &\textbf{78.25} & \textbf{84.83} & \textbf{90.17}& \textbf{85.78} \\
            \bottomrule
        \end{tabular}
    }
    \vspace{-0.3cm}
\end{table*}
\subsubsection{Evaluation on Zero-Shot ECG Classification}
We evaluate \modelname's zero-shot classification performance against the multimodal baseline MERL~\cite{liu2024merl}. 
As shown in Table~\ref{tab:zero_shot}, \modelname\ achieves state-of-the-art results across all tasks, demonstrating significant performance improvements.
Specifically, \modelname\ demonstrates improvements of $+3.2\%$, $+6.9\%$, $+3.2\%$, $+5.5\%$, $+2.0\%$, $+1.4\%$ on CSN, PTBXL-Rhythm, PTBXL-Form, PTBXL-Sub, PTBXL-Super, and CPSC2018 respectively.
Moreover, \modelname\ achieves an average performance gain of 3.7\% over the MERL.
These findings highlight \modelname's effectiveness in zero-shot ECG classification, reinforcing its potential for clinical applications where evaluation on diverse downstream tasks is required without additional training.

\subsubsection{Evaluation on Transfer Learning for ECG Classification}
To assess robustness to domain shifts, we evaluate transfer learning performance using test datasets that differ in distribution from the pretraining data but share the same label space. For non-zero-shot self-supervised learning (SSL) baselines, we apply linear probing using 100\% of the source domain training data. 
For zero-shot models (MERL), we follow the target-source category matching protocol outlined in Appendix~\ref{sec:pretraining_details}. This evaluation measures \modelname's ability to generalize across diverse clinical settings.
As shown in Table~\ref{tab:transfer}, \modelname\ achieves the highest performance in three out of six settings and ranks second in two. Notably, it also achieves the best average performance across all six tasks. 
Specifically,  our \modelname\ outperforms MERL by $+1.12\%$, $+3.76\%$, $+0.52\%$ on the CPSC2018 $\rightarrow$ PTXBL-Super, CPSC2018 $\rightarrow$ CSN and CSN $\rightarrow$ PTXBL-Super settings, respectively.
These results highlights \modelname's robustness and its ability to handle distribution shifts effectively.
\subsection{Analysis of Our Framework}

\begin{table}[t!]
    \centering
    \caption{
    ECG report generation metrics ([\%]) on 500 curated samples from PTB-XL report dataset.
    }
    \label{tab:report_generation_results}
    \resizebox{\linewidth}{!}{
        \begin{tabular}{c | c | c c c c c c c}
            \toprule
            Models & Size & BLEU-1 & BLEU-4 & METEOR & ROUGE-L & BERTScore F1 \\
            \midrule
            PULSE~\cite{liu2024teach} & 7B & 5.12 & 0.83 & \textbf{13.76} & 8.15 & 10.96 \\
            \modelname\ & 284M & \textbf{13.02} & \textbf{1.87} & 11.28 & \textbf{18.50} & \textbf{44.08} \\
            \bottomrule
        \end{tabular}
    }
    \vspace{-0.6cm}
\end{table}
\subsubsection{Ablation on Multi-scale Supervision}  
To evaluate the impact of multi-scale supervision, including token-level supervision ($\mathcal{L}_{\mathrm{LM}}$), beat-level supervision ($\mathcal{L}_{\mathrm{Local}}$), and rhythm-level supervision ($\mathcal{L}_{\mathrm{g}}$), we conduct ablation studies over different variants. This allows us to assess each supervision level’s contribution.
As shown in Table~\ref{tab:ablation_of_loss}, 
The first three rows present models trained with single isolated supervision (one loss each). Rows four and five correspond to variants without token-level supervision ($\mathcal{L}_{\mathrm{LM}}$) and without beat-level supervision ($\mathcal{L}_{\mathrm{Local}}$), respectively. The final row shows the performance of the full model. Our full model achieves the highest performance on most benchmarks, outperforming the best partial-supervision variant by an average AUROC gain of $2.86\%$.
Furthermore, we observe that variants without rhythm-level supervision ($\mathcal{L}{\mathrm{g}}$) perform significantly worse, while all configurations including $\mathcal{L}{\mathrm{g}}$ achieve AUROCs above 82\%. This demonstrates that global-level multi-modal alignment is essential for effective ECG representation learning.
This empirically demonstrates the importance of hierarchical supervision—integrating rhythm-level ($\mathcal{L}_{\mathrm{g}}$), beat-level ($\mathcal{L}_{\mathrm{Local}}$), and token-level ($\mathcal{L}_{\mathrm{LM}}$) signals—for learning clinically meaningful ECG representations.
Furthermore, we extend this analysis to zero-shot classification following the framework used in previous ablation studies. As shown in Table~\ref{tab:ablation_zero_shot}, the full model consistently outperforms ablated versions, particularly in low-data settings, with an average AUROC improvement of 2.1\%. These findings highlight the critical role of multi-scale supervision in learning robust and transferable ECG representations.
\subsubsection{Analysis on ECG Report Generation}
Although \modelname\ is primarily designed to encode generalized global ECG representations, we further evaluate its fine-grained prediction capabilities to validate the effectiveness of our multi-scale pretraining approach.
Specifically, we conduct a preliminary evaluation on the ECG report generation task using the ECGBench dataset~\cite{liu2024teach}, and compare our model with PULSE~\cite{liu2024teach}, a multimodal baseline trained on ECG images via instruction tuning and evaluated in a zero-shot setting.
We employ both standard Natural Language Generation (NLG) metrics and BERTScore to assess the lexical and semantic quality of the generated reports.
As shown in Table~\ref{tab:report_generation_results}, \modelname\ significantly outperforms PULSE, demonstrating strong capabilities in fine-grained ECG understanding and report generation.
These results highlight the potential of \modelname\ to support clinically relevant diagnostic tasks that require detailed, fine-grained predictions.
Additionally, we evaluate the model on patient identification, with results reported in Table~\ref{tab:patient_identification_results}.
A more comprehensive analysis of fine-grained prediction performance remains an important direction for future research.

\subsubsection{Ablation on ECG Encoder}
\label{sec:ablation on ecg encoder}
Table~\ref{tab:ablation_of_ecg_encoder} presents an ablation study evaluating the impact of different ECG encoder backbones on the pretraining performance of \modelname.
Specifically, we compare the Wav2Vec 2.0~\cite{baevski2020wav2vec} architecture with the ResNet-18 backbone used in MERL~\cite{liu2024merl}. In addition, we assess the effectiveness of using CMSC~\cite{kiyasseh2021clocs} for pretraining the ECG encoder prior to multimodal training.
CMSC is a patient-specific contrastive learning method for unlabeled ECG data and is orthogonal to the choice of ECG encoder architecture.
The results show that employing Wav2Vec 2.0 as the ECG encoder consistently outperforms the CMSC-based variant across all evaluation settings, with an average gain of $+3.53\%$. Furthermore, it outperforms the ResNet-18-based variant in 17 out of 18 settings, with an average improvement of $+3.63\%$.
%
% Thus, we use Wav2Vec 2.0 as the ECG encoder backbone by default.
%
\begin{table*}[t!]
    \centering
    \caption{Ablation results of hyperparameters on 6 zero-shot classification tasks.}
    \label{tab:ablation_hyperparameter}
    \resizebox{0.75\textwidth}{!}{
    \begin{tabular}{c c | c c c c c c | c}
        \toprule
        $\lambda_1$ & $\lambda_2$ & PTBXL-Rhythm  & PTBXL-Form  & PTBXL-Sub &  
           PTBXL-Super & CPSC2018 & CSN & Average  \\
        \midrule
        1 & 1 & 85.3 & 66.3 & 76.9 & 75.3 & 82.9 & 75.6 & 77.1 \\
        1 & 0.2 & 84.6 & 69.0 & 75.2 & 76.1 & 84.0 & 77.0 & 77.6 \\
        2 & 1 & 79.6 & 68.7 & 77.1 & 77.2 & 83.9 & 75.2 & 76.9 \\
        2 & 0.2 & \textbf{85.4} & \textbf{69.1} & \textbf{81.2} & \textbf{76.2} & \textbf{84.2} & \textbf{77.6} & \textbf{78.9} \\
        \bottomrule
    \end{tabular}
    }
    \vspace{-0.4cm}
\end{table*}
\subsubsection{Ablation on Uni-modal Text Pretraining}  
To evaluate the impact of uni-modal text pretraining, we compare the full model with a variant that omits this step. As shown in Table~\ref{tab:ablation_of_text}, the baseline model (first row) operates without text pretraining, while the full model (second row) incorporates it. The full model outperforms the baseline in 13 out of 18 settings, achieving an average improvement of 0.35\%.
These results suggest that text pretraining enhances downstream generalization by improving feature separation in the joint embedding space. They also highlight the potential benefits of developing more specialized cardiology-specific language models to further improve performance.
\subsubsection{Ablation on Hyperparameters}
\label{sec:ablation_hyperparameters}
We evaluate the impact of hyperparameters $\lambda_1$ and $\lambda_2$ on zero-shot classification performance, as shown in Table~\ref{tab:ablation_hyperparameter}.
Overall, the performance remains consistent across four different hyperparameter configurations, demonstrating the robustness of our framework. Through a preliminary search, we identify $\lambda_1 = 2$ and $\lambda_2 = 0.2 $ as the optimal setting, which we adopt as the default.

\subsection{Qualitative Results}
\begin{figure}[t!]
    \centering
    \includegraphics[width=\linewidth]{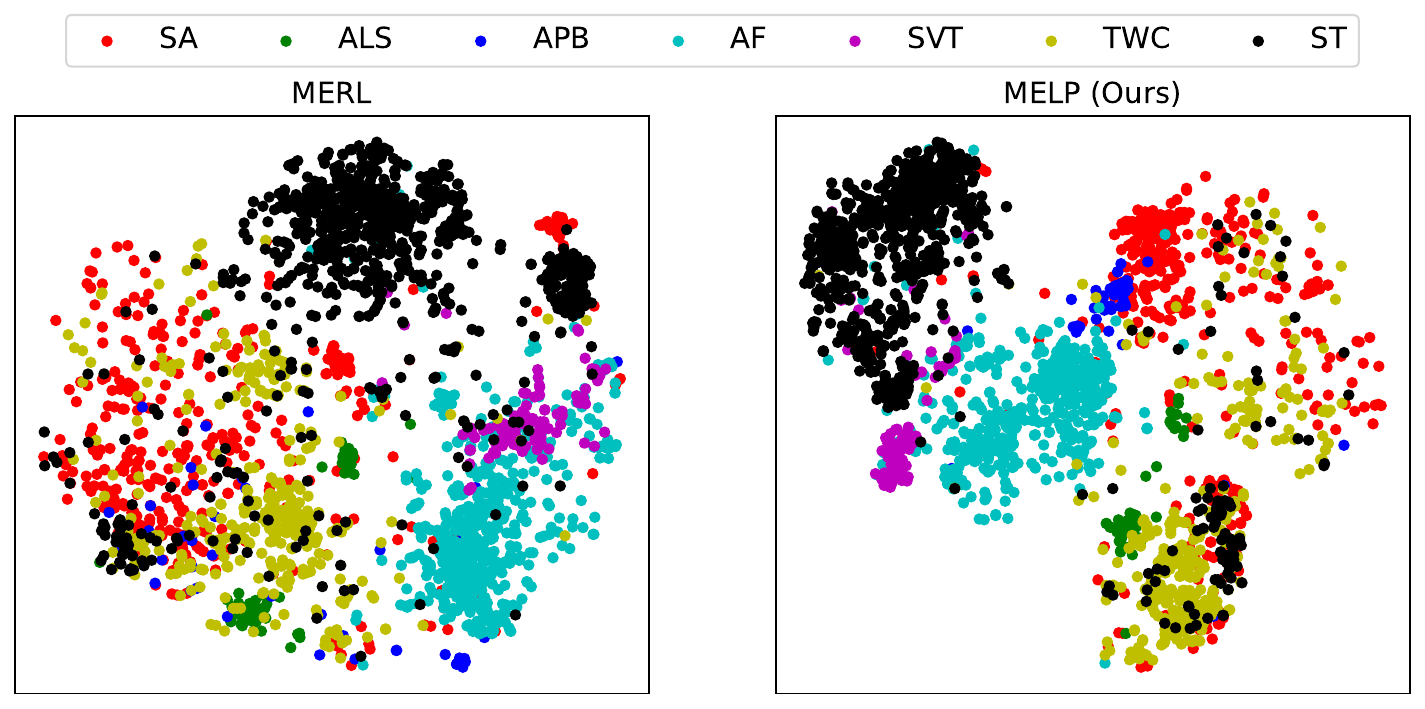}
    \vspace{-0.4cm}
    \caption{Comparison of T-SNE visualizations of the embedding space for MERL and \modelname\ on the CSN test set. }
    \label{fig:tsne}
    \vspace{-0.5cm}
\end{figure}

To analyze the learned representations, we visualize the embedding space of the CSN test set in Figure~\ref{fig:tsne}. Following~\cite{liu2024merl}, we focus on seven common ECG abnormalities and select samples that exclusively exhibit each condition.
We extract embeddings using both MERL and \modelname\ and project them into a lower-dimensional space for visualization. The results show that \modelname\ produces a more distinct and well-separated embedding space compared to MERL. This improved separability helps explain \modelname's superior zero-shot classification performance.
%
% \section{Conclusion}

% In conclusion, we propose a multi-modal ECG foundation model with multi-scale supervision from clinical text report to learn fine-grained representations for various downstream tasks. We run experiments on six downstream datasets and show our model's outstanding performance compared with other baseline models.

% \para{Limitations and Future Work.}

% Though our method, MLEP, learn representations from token, beats and rhythm levels and transfer knowledge from language modality, it is difficult for our model to provide interpretation and clinical explanation for token level supervision. Meanwhile, we simply use 10 learnable queries through the attention pooler to aggregate the beats level representations, which may not be accurate for some cases where one ECG recordings contains more than 10 beats. Therefore, future work may focus on seperate the ECG recordings based on more accurate and instructive clinical supervisions.  

\section{Conclusion}  
We introduce \modelname, a multimodal ECG foundation model that leverages multi-scale supervision from clinical text reports to learn fine-grained representations for diverse downstream tasks. Extensive experiments on three benchmark datasets demonstrate its superior performance over existing baselines, highlighting its effectiveness in aligning ECG signals with clinical text at multiple levels of abstraction.

\para{Limitations and Future Directions.}
Despite its strengths, \modelname\ has two limitations. First, its token-level supervision lacks explicit clinical interpretability, limiting its applicability in explainable diagnostic settings, as discussed in Appendix~\ref{sec:multi-level-design}.
Future work will explore leveraging external medical knowledge bases to generate clinically meaningful descriptions, thereby enabling more interpretable and informative token-level supervision.
Second, the current model adopts the MedCPT architecture as its text encoder, which does not fully capitalize on the recent advances in large language models (LLMs).
To address this, future research will investigate ECG instruction tuning frameworks that harness the strong generalization and reasoning capabilities of LLMs for enhanced multimodal understanding.
\section*{Acknowledgment}
This work was supported in part by the Research Grants Council of Hong Kong (27206123, C5055-24G, and T45-401/22-N), the Hong Kong Innovation and Technology Fund (ITS/273/22, ITS/274/22, and GHP/318/22GD), the National Natural Science Foundation of China (No. 62201483), and Guangdong Natural Science Fund (No. 2024A1515011875).

\section*{Impact Statement}

The pretraining of \modelname\ is based on large-scale ECG-text pairs. In this study, we utilize the publicly available MIMIC-IV-ECG dataset, which has undergone rigorous de-identification procedures. Nonetheless, privacy considerations remain critical, especially when deploying the framework in other scenarios.
Moreover, although \modelname\ does not explicitly incorporate sensitive attributes such as gender or race during pretraining, its predictions may still reflect subtle biases related to these factors.
We encourage future work to further investigate privacy implications and develop effective bias-mitigation strategies to ensure the ethical and equitable deployment of AI systems in healthcare.
In addition, adapting our framework for clinical scenarios will require careful consideration of regulatory requirements.


\clearpage
\bibliography{example_paper}

\begin{thebibliography}{51}
\providecommand{\natexlab}[1]{#1}
\providecommand{\url}[1]{\texttt{#1}}
\expandafter\ifx\csname urlstyle\endcsname\relax
  \providecommand{\doi}[1]{doi: #1}\else
  \providecommand{\doi}{doi: \begingroup \urlstyle{rm}\Url}\fi

\bibitem[Baevski et~al.(2020)Baevski, Zhou, Mohamed, and Auli]{baevski2020wav2vec}
Baevski, A., Zhou, Y., Mohamed, A., and Auli, M.
\newblock wav2vec 2.0: A framework for self-supervised learning of speech representations.
\newblock \emph{Advances in neural information processing systems}, 33:\penalty0 12449--12460, 2020.

\bibitem[Boecking et~al.(2022)Boecking, Usuyama, Bannur, Castro, Schwaighofer, Hyland, Wetscherek, Naumann, Nori, Alvarez-Valle, et~al.]{boecking2022making}
Boecking, B., Usuyama, N., Bannur, S., Castro, D.~C., Schwaighofer, A., Hyland, S., Wetscherek, M., Naumann, T., Nori, A., Alvarez-Valle, J., et~al.
\newblock Making the most of text semantics to improve biomedical vision--language processing.
\newblock In \emph{European conference on computer vision}, pp.\  1--21. Springer, 2022.

\bibitem[Caron et~al.(2021)Caron, Touvron, Misra, J{\'e}gou, Mairal, Bojanowski, and Joulin]{caron2021emerging}
Caron, M., Touvron, H., Misra, I., J{\'e}gou, H., Mairal, J., Bojanowski, P., and Joulin, A.
\newblock Emerging properties in self-supervised vision transformers.
\newblock In \emph{Proceedings of the IEEE/CVF international conference on computer vision}, pp.\  9650--9660, 2021.

\bibitem[Chen et~al.(2020)Chen, Kornblith, Norouzi, and Hinton]{chen2020simple}
Chen, T., Kornblith, S., Norouzi, M., and Hinton, G.
\newblock A simple framework for contrastive learning of visual representations.
\newblock In \emph{International conference on machine learning}, pp.\  1597--1607. PMLR, 2020.

\bibitem[Chen \& He(2021)Chen and He]{chen2021exploring}
Chen, X. and He, K.
\newblock Exploring simple siamese representation learning.
\newblock In \emph{Proceedings of the IEEE/CVF conference on computer vision and pattern recognition}, pp.\  15750--15758, 2021.

\bibitem[Chen et~al.(2021)Chen, Xie, and He]{chen2021empirical}
Chen, X., Xie, S., and He, K.
\newblock An empirical study of training self-supervised vision transformers.
\newblock In \emph{Proceedings of the IEEE/CVF international conference on computer vision}, pp.\  9640--9649, 2021.

\bibitem[Devlin(2018)]{devlin2018bert}
Devlin, J.
\newblock Bert: Pre-training of deep bidirectional transformers for language understanding.
\newblock \emph{arXiv preprint arXiv:1810.04805}, 2018.

\bibitem[Dong et~al.(2019)Dong, Yang, Wang, Wei, Liu, Wang, Gao, Zhou, and Hon]{dong2019unified}
Dong, L., Yang, N., Wang, W., Wei, F., Liu, X., Wang, Y., Gao, J., Zhou, M., and Hon, H.-W.
\newblock Unified language model pre-training for natural language understanding and generation.
\newblock \emph{Advances in neural information processing systems}, 32, 2019.

\bibitem[Ebrahimi et~al.(2020)Ebrahimi, Loni, Daneshtalab, and Gharehbaghi]{ebrahimi2020review}
Ebrahimi, Z., Loni, M., Daneshtalab, M., and Gharehbaghi, A.
\newblock A review on deep learning methods for ecg arrhythmia classification.
\newblock \emph{Expert Systems with Applications: X}, 7:\penalty0 100033, 2020.

\bibitem[Eldele et~al.(2021)Eldele, Ragab, Chen, Wu, Kwoh, Li, and Guan]{eldele2021time}
Eldele, E., Ragab, M., Chen, Z., Wu, M., Kwoh, C.~K., Li, X., and Guan, C.
\newblock Time-series representation learning via temporal and contextual contrasting.
\newblock \emph{arXiv preprint arXiv:2106.14112}, 2021.

\bibitem[Gopal et~al.(2021)Gopal, Han, Raghupathi, Ng, Tison, and Rajpurkar]{gopal20213kg}
Gopal, B., Han, R., Raghupathi, G., Ng, A., Tison, G., and Rajpurkar, P.
\newblock 3kg: Contrastive learning of 12-lead electrocardiograms using physiologically-inspired augmentations.
\newblock In \emph{Machine Learning for Health}, pp.\  156--167. PMLR, 2021.

\bibitem[Gow et~al.(2023)Gow, Pollard, Nathanson, Johnson, Moody, Fernandes, Greenbaum, Berkowitz, Moukheiber, Eslami, et~al.]{gow2023mimic}
Gow, B., Pollard, T., Nathanson, L.~A., Johnson, A., Moody, B., Fernandes, C., Greenbaum, N., Berkowitz, S., Moukheiber, D., Eslami, P., et~al.
\newblock Mimic-iv-ecg-diagnostic electrocardiogram matched subset.
\newblock \emph{Type: dataset}, 2023.

\bibitem[Grill et~al.(2020)Grill, Strub, Altch{\'e}, Tallec, Richemond, Buchatskaya, Doersch, Avila~Pires, Guo, Gheshlaghi~Azar, et~al.]{grill2020bootstrap}
Grill, J.-B., Strub, F., Altch{\'e}, F., Tallec, C., Richemond, P., Buchatskaya, E., Doersch, C., Avila~Pires, B., Guo, Z., Gheshlaghi~Azar, M., et~al.
\newblock Bootstrap your own latent-a new approach to self-supervised learning.
\newblock \emph{Advances in neural information processing systems}, 33:\penalty0 21271--21284, 2020.

\bibitem[Gwon et~al.(2024)Gwon, Seo, Park, Kim, and Jun]{gwon2024medical}
Gwon, H., Seo, J., Park, S., Kim, Y.-H., and Jun, T.~J.
\newblock Medical language model specialized in extracting cardiac knowledge.
\newblock \emph{Scientific Reports}, 14\penalty0 (1):\penalty0 29059, 2024.

\bibitem[Han et~al.(2024)Han, Liu, Zhang, and Ding]{han2024foundation}
Han, Y., Liu, X., Zhang, X., and Ding, C.
\newblock Foundation models in electrocardiogram: A review.
\newblock \emph{arXiv preprint arXiv:2410.19877}, 2024.

\bibitem[He et~al.(2020{\natexlab{a}})He, Fan, Wu, Xie, and Girshick]{he2020momentum}
He, K., Fan, H., Wu, Y., Xie, S., and Girshick, R.
\newblock Momentum contrast for unsupervised visual representation learning.
\newblock In \emph{Proceedings of the IEEE/CVF conference on computer vision and pattern recognition}, pp.\  9729--9738, 2020{\natexlab{a}}.

\bibitem[He et~al.(2020{\natexlab{b}})He, Liu, Gao, and Chen]{he2020deberta}
He, P., Liu, X., Gao, J., and Chen, W.
\newblock Deberta: Decoding-enhanced bert with disentangled attention.
\newblock \emph{arXiv preprint arXiv:2006.03654}, 2020{\natexlab{b}}.

\bibitem[Hu et~al.(2023)Hu, Chen, and Zhou]{hu2023spatiotemporal}
Hu, R., Chen, J., and Zhou, L.
\newblock Spatiotemporal self-supervised representation learning from multi-lead ecg signals.
\newblock \emph{Biomedical Signal Processing and Control}, 84:\penalty0 104772, 2023.

\bibitem[Jin et~al.(2025)Jin, Wang, Li, Li, Pan, and Hong]{jin2025reading}
Jin, J., Wang, H., Li, H., Li, J., Pan, J., and Hong, S.
\newblock Reading your heart: Learning {ECG} words and sentences via pre-training {ECG} language model.
\newblock In \emph{The Thirteenth International Conference on Learning Representations}, 2025.
\newblock URL \url{https://openreview.net/forum?id=6Hz1Ko087B}.

\bibitem[Jin et~al.(2023)Jin, Kim, Chen, Comeau, Yeganova, Wilbur, and Lu]{jin2023medcpt}
Jin, Q., Kim, W., Chen, Q., Comeau, D.~C., Yeganova, L., Wilbur, W.~J., and Lu, Z.
\newblock Medcpt: Contrastive pre-trained transformers with large-scale pubmed search logs for zero-shot biomedical information retrieval.
\newblock \emph{Bioinformatics}, 39\penalty0 (11):\penalty0 btad651, 2023.

\bibitem[Kiyasseh et~al.(2021)Kiyasseh, Zhu, and Clifton]{kiyasseh2021clocs}
Kiyasseh, D., Zhu, T., and Clifton, D.~A.
\newblock Clocs: Contrastive learning of cardiac signals across space, time, and patients.
\newblock In \emph{International Conference on Machine Learning}, pp.\  5606--5615. PMLR, 2021.

\bibitem[LB(2011)]{lb2011ccs}
LB, M.
\newblock Ccs atrial fibrillation guidelines committee: Canadian cardiovascular society atrial fibrillation guidelines 2010: Prevention and treatment of atrial fibrillation following cardiac surgery.
\newblock \emph{Can J Cardiol}, 27:\penalty0 91--97, 2011.

\bibitem[Li et~al.(2024)Li, Liu, Cheng, Arcucci, and Hong]{li2024frozen}
Li, J., Liu, C., Cheng, S., Arcucci, R., and Hong, S.
\newblock Frozen language model helps ecg zero-shot learning.
\newblock In \emph{Medical Imaging with Deep Learning}, pp.\  402--415. PMLR, 2024.

\bibitem[Liu et~al.(2024{\natexlab{a}})Liu, Wan, Ouyang, Shah, Bai, and Arcucci]{liu2024merl}
Liu, C., Wan, Z., Ouyang, C., Shah, A., Bai, W., and Arcucci, R.
\newblock Zero-shot ecg classification with multimodal learning and test-time clinical knowledge enhancement.
\newblock \emph{arXiv preprint arXiv:2403.06659}, 2024{\natexlab{a}}.

\bibitem[Liu et~al.(2018)Liu, Liu, Zhao, Zhang, Wu, Xu, Liu, Ma, Wei, He, et~al.]{liu2018open}
Liu, F., Liu, C., Zhao, L., Zhang, X., Wu, X., Xu, X., Liu, Y., Ma, C., Wei, S., He, Z., et~al.
\newblock An open access database for evaluating the algorithms of electrocardiogram rhythm and morphology abnormality detection.
\newblock \emph{Journal of Medical Imaging and Health Informatics}, 8\penalty0 (7):\penalty0 1368--1373, 2018.

\bibitem[Liu et~al.(2024{\natexlab{b}})Liu, Bai, Yue, and Zhang]{liu2024teach}
Liu, R., Bai, Y., Yue, X., and Zhang, P.
\newblock Teach multimodal llms to comprehend electrocardiographic images.
\newblock \emph{arXiv preprint arXiv:2410.19008}, 2024{\natexlab{b}}.

\bibitem[Mattu et~al.(2019)Mattu, Tabas, and Brady]{mattu2019electrocardiography}
Mattu, A., Tabas, J.~A., and Brady, W.~J.
\newblock \emph{Electrocardiography in emergency, acute, and critical care}.
\newblock American College of Emergency Physicians, 2019.

\bibitem[McKeen et~al.(2024)McKeen, Oliva, Masood, Toma, Rubin, and Wang]{mckeen2024ecg}
McKeen, K., Oliva, L., Masood, S., Toma, A., Rubin, B., and Wang, B.
\newblock Ecg-fm: An open electrocardiogram foundation model.
\newblock \emph{arXiv preprint arXiv:2408.05178}, 2024.

\bibitem[Na et~al.(2024)Na, Park, Tae, and Joo]{na2024guiding}
Na, Y., Park, M., Tae, Y., and Joo, S.
\newblock Guiding masked representation learning to capture spatio-temporal relationship of electrocardiogram.
\newblock \emph{arXiv preprint arXiv:2402.09450}, 2024.

\bibitem[Nie et~al.(2022)Nie, Nguyen, Sinthong, and Kalagnanam]{nie2022time}
Nie, Y., Nguyen, N.~H., Sinthong, P., and Kalagnanam, J.
\newblock A time series is worth 64 words: Long-term forecasting with transformers.
\newblock \emph{arXiv preprint arXiv:2211.14730}, 2022.

\bibitem[Oh et~al.(2022)Oh, Chung, Kwon, Hong, and Choi]{oh2022lead}
Oh, J., Chung, H., Kwon, J.-m., Hong, D.-g., and Choi, E.
\newblock Lead-agnostic self-supervised learning for local and global representations of electrocardiogram.
\newblock In \emph{Conference on Health, Inference, and Learning}, pp.\  338--353. PMLR, 2022.

\bibitem[Oord et~al.(2018)Oord, Li, and Vinyals]{oord2018representation}
Oord, A. v.~d., Li, Y., and Vinyals, O.
\newblock Representation learning with contrastive predictive coding.
\newblock \emph{arXiv preprint arXiv:1807.03748}, 2018.

\bibitem[Pham et~al.(2024)Pham, Saeed, and Ma]{pham2024cmelt}
Pham, M., Saeed, A., and Ma, D.
\newblock C-melt: Contrastive enhanced masked auto-encoders for ecg-language pre-training.
\newblock \emph{arXiv preprint arXiv:2410.02131}, 2024.

\bibitem[Radford et~al.(2021)Radford, Kim, Hallacy, Ramesh, Goh, Agarwal, Sastry, Askell, Mishkin, Clark, et~al.]{radford2021learning}
Radford, A., Kim, J.~W., Hallacy, C., Ramesh, A., Goh, G., Agarwal, S., Sastry, G., Askell, A., Mishkin, P., Clark, J., et~al.
\newblock Learning transferable visual models from natural language supervision.
\newblock In \emph{International conference on machine learning}, pp.\  8748--8763. PMLR, 2021.

\bibitem[Sangha et~al.(2024)Sangha, Khunte, Holste, Mortazavi, Wang, Oikonomou, and Khera]{sangha2024biometric}
Sangha, V., Khunte, A., Holste, G., Mortazavi, B.~J., Wang, Z., Oikonomou, E.~K., and Khera, R.
\newblock Biometric contrastive learning for data-efficient deep learning from electrocardiographic images.
\newblock \emph{Journal of the American Medical Informatics Association}, 31\penalty0 (4):\penalty0 855--865, 2024.

\bibitem[Siontis et~al.(2021)Siontis, Noseworthy, Attia, and Friedman]{siontis2021artificial}
Siontis, K.~C., Noseworthy, P.~A., Attia, Z.~I., and Friedman, P.~A.
\newblock Artificial intelligence-enhanced electrocardiography in cardiovascular disease management.
\newblock \emph{Nature Reviews Cardiology}, 18\penalty0 (7):\penalty0 465--478, 2021.

\bibitem[Song et~al.(2024)Song, Jang, Lee, Hong, Kwon, and Jo]{song2024foundation}
Song, J., Jang, J.-H., Lee, B.~T., Hong, D., Kwon, J.-m., and Jo, Y.-Y.
\newblock Foundation models for ecg: Leveraging hybrid self-supervised learning for advanced cardiac diagnostics.
\newblock \emph{arXiv preprint arXiv:2407.07110}, 2024.

\bibitem[Tian et~al.(2024)Tian, Li, Jin, Wang, Wei, Zhao, Liu, Liu, and Liu]{tian2024foundation}
Tian, Y., Li, Z., Jin, Y., Wang, M., Wei, X., Zhao, L., Liu, Y., Liu, J., and Liu, C.
\newblock Foundation model of ecg diagnosis: Diagnostics and explanations of any form and rhythm on ecg.
\newblock \emph{Cell Reports Medicine}, 5\penalty0 (12), 2024.

\bibitem[Wagner et~al.(2020)Wagner, Strodthoff, Bousseljot, Kreiseler, Lunze, Samek, and Schaeffter]{wagner2020ptb}
Wagner, P., Strodthoff, N., Bousseljot, R.-D., Kreiseler, D., Lunze, F.~I., Samek, W., and Schaeffter, T.
\newblock Ptb-xl, a large publicly available electrocardiography dataset.
\newblock \emph{Scientific data}, 7\penalty0 (1):\penalty0 1--15, 2020.

\bibitem[Wang et~al.(2024)Wang, Feng, Ge, Zhou, Zhou, and Wang]{10177892}
Wang, N., Feng, P., Ge, Z., Zhou, Y., Zhou, B., and Wang, Z.
\newblock Adversarial spatiotemporal contrastive learning for electrocardiogram signals.
\newblock \emph{IEEE Transactions on Neural Networks and Learning Systems}, 35\penalty0 (10):\penalty0 13845--13859, 2024.
\newblock \doi{10.1109/TNNLS.2023.3272153}.

\bibitem[Williams \& Zipser(1989)Williams and Zipser]{williams1989learning}
Williams, R.~J. and Zipser, D.
\newblock A learning algorithm for continually running fully recurrent neural networks.
\newblock \emph{Neural computation}, 1\penalty0 (2):\penalty0 270--280, 1989.

\bibitem[Wu et~al.(2018)Wu, Xiong, Yu, and Lin]{wu2018unsupervised}
Wu, Z., Xiong, Y., Yu, S.~X., and Lin, D.
\newblock Unsupervised feature learning via non-parametric instance discrimination.
\newblock In \emph{Proceedings of the IEEE conference on computer vision and pattern recognition}, pp.\  3733--3742, 2018.

\bibitem[Yan et~al.(2019)Yan, Liang, Zhang, and Liu]{yan2019fusing}
Yan, G., Liang, S., Zhang, Y., and Liu, F.
\newblock Fusing transformer model with temporal features for ecg heartbeat classification.
\newblock In \emph{2019 IEEE International Conference on Bioinformatics and Biomedicine (BIBM)}, pp.\  898--905. IEEE, 2019.

\bibitem[Yu et~al.(2023)Yu, Yang, and Sano]{yu2023ecg}
Yu, H., Yang, H., and Sano, A.
\newblock Ecg-sl: Electrocardiogram (ecg) segment learning, a deep learning method for ecg signal.
\newblock \emph{arXiv preprint arXiv:2310.00818}, 2023.

\bibitem[Yu et~al.(2024)Yu, Guo, and Sano]{yu2024ecg}
Yu, H., Guo, P., and Sano, A.
\newblock Ecg semantic integrator (esi): A foundation ecg model pretrained with llm-enhanced cardiological text.
\newblock \emph{arXiv preprint arXiv:2405.19366}, 2024.

\bibitem[Yu et~al.(2022)Yu, Wang, Vasudevan, Yeung, Seyedhosseini, and Wu]{yu2022coca}
Yu, J., Wang, Z., Vasudevan, V., Yeung, L., Seyedhosseini, M., and Wu, Y.
\newblock Coca: Contrastive captioners are image-text foundation models.
\newblock \emph{arXiv preprint arXiv:2205.01917}, 2022.

\bibitem[Yue et~al.(2022)Yue, Wang, Duan, Yang, Huang, Tong, and Xu]{yue2022ts2vec}
Yue, Z., Wang, Y., Duan, J., Yang, T., Huang, C., Tong, Y., and Xu, B.
\newblock Ts2vec: Towards universal representation of time series.
\newblock In \emph{Proceedings of the AAAI Conference on Artificial Intelligence}, volume~36, pp.\  8980--8987, 2022.

\bibitem[Zbontar et~al.(2021)Zbontar, Jing, Misra, LeCun, and Deny]{zbontar2021barlow}
Zbontar, J., Jing, L., Misra, I., LeCun, Y., and Deny, S.
\newblock Barlow twins: Self-supervised learning via redundancy reduction.
\newblock In \emph{International conference on machine learning}, pp.\  12310--12320. PMLR, 2021.

\bibitem[Zhang et~al.(2023)Zhang, Yang, Geng, and Hong]{zhang2023self}
Zhang, W., Yang, L., Geng, S., and Hong, S.
\newblock Self-supervised time series representation learning via cross reconstruction transformer.
\newblock \emph{IEEE Transactions on Neural Networks and Learning Systems}, 2023.

\bibitem[Zhao et~al.(2024)Zhao, Zhang, Wang, Han, Chen, Huang, Jin, and Kang]{zhao2024ecg}
Zhao, Y., Zhang, T., Wang, X., Han, P., Chen, T., Huang, L., Jin, Y., and Kang, J.
\newblock Ecg-chat: A large ecg-language model for cardiac disease diagnosis.
\newblock \emph{arXiv preprint arXiv:2408.08849}, 2024.

\bibitem[Zheng et~al.(2022)Zheng, Guo, and Chu]{zheng2022large}
Zheng, J., Guo, H., and Chu, H.
\newblock A large scale 12-lead electrocardiogram database for arrhythmia study (version 1.0. 0).
\newblock \emph{PhysioNet 2022Available online httpphysionet orgcontentecg arrhythmia10 0accessed on}, 23, 2022.

\end{thebibliography}
\bibliographystyle{icml2025}

%%%%%%%%%%%%%%%%%%%%%%%%%%%%%%%%%%%%%%%%%%%%%%%%%%%%%%%%%%%%%%%%%%%%%%%%%%%%%%%
%%%%%%%%%%%%%%%%%%%%%%%%%%%%%%%%%%%%%%%%%%%%%%%%%%%%%%%%%%%%%%%%%%%%%%%%%%%%%%%
% APPENDIX
%%%%%%%%%%%%%%%%%%%%%%%%%%%%%%%%%%%%%%%%%%%%%%%%%%%%%%%%%%%%%%%%%%%%%%%%%%%%%%%
%%%%%%%%%%%%%%%%%%%%%%%%%%%%%%%%%%%%%%%%%%%%%%%%%%%%%%%%%%%%%%%%%%%%%%%%%%%%%%%
\newpage
\appendix
\onecolumn
\section{Details of Training Cardiology Language Model}
\label{sec:cardiology_language}

\subsection{Cardiology Corpus Details}
\label{sec:cardiology_corpus}
To extract cardiology-related data from PubMed and Wikipedia, we followed the procedure used in HeartBERT~\cite{gwon2024medical}. 
For PudMed dataset, we compiled a list of cardiology-related journal names from the SJR (Scimago Journal \& Country Rank) database, along with glossaries from Aiken, NIH, and the Texas Heart Institute. These terms were used as queries to retrieve relevant content via the PudMed database API. To ensure content relevance, we only employ abstract section for PudMed dataset.
For Wikipedia dataset, since it already provides information about categories and subcategories for classification, we use a top-level category called “Cardiology” as the primary category. Starting with the “Cardiology” category, we navigated through the subcategories provided by Wikipedia to collect related articles. 
To ensure content relevance, we only employ abstract section for PudMed dataset.
This process finally resulted in a curated dataset of approximately 5.6 GB, containing 912.5 million corpus.
In addition to the curated corpus introduced in Section~\ref{sec:cardiology_corpus}, we further incorporate ECG-related reports from the MIMIC-IV-ECG training set to enhance the model’s understanding of ECG-specific clinical language.

\subsection{Training Details}
\begin{figure*}[ht!]
    \centering
    \includegraphics[width=0.7\textwidth]{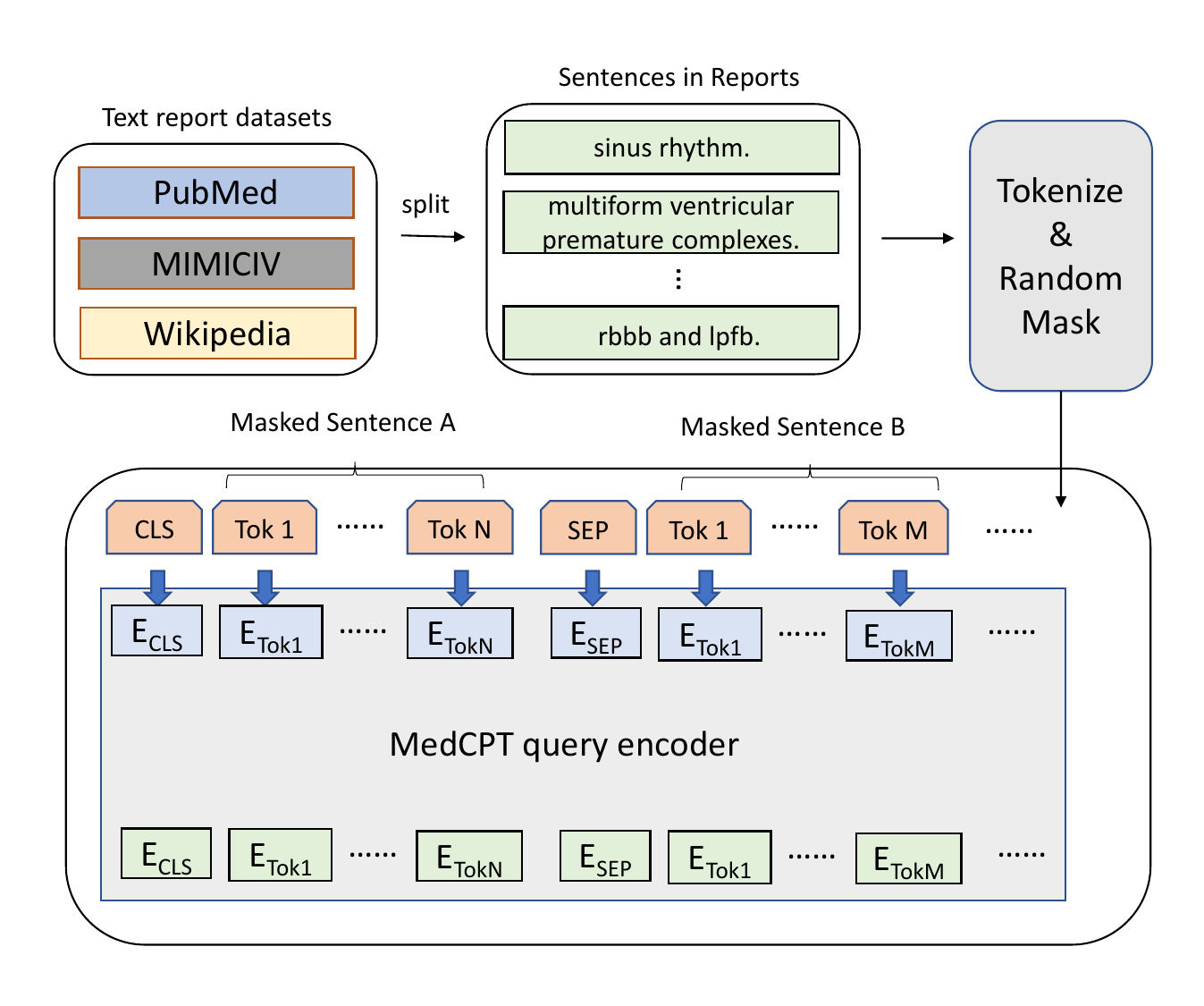}
    \vspace{-0.5cm}
    \caption{Framework for text uni-modal pretraining. Three widely-used medical datasets were combined and segmented into sentences as model inputs. Following BERT-style masking methodology~\cite{devlin2018bert}, we randomly mask portions of the input text and insert special tokens between sentences. The masked sequences are then processed through the MedCPT query encoder to generate text representations.}
    \label{fig:text-encoder}
\end{figure*}

Figure~\ref{fig:text-encoder} illustrates the overall training workflow of our cardiology-specific language model pretraining.
We initialize our model using the query encoder from MedCPT~\cite{jin2023medcpt}\footnote{https://huggingface.co/ncbi/MedCPT-Query-Encoder}, which was originally trained on PubMed search logs. To better adapt it to the cardiology domain, we further fine-tune this encoder using a masked language modeling (MLM) objective.
Specifically, we apply random masking to tokens in ECG diagnostic reports and train the model to reconstruct the masked tokens based on their surrounding context. This encourages the encoder to capture nuanced, domain-specific semantics from cardiology narratives.
\vspace{-0.4cm}
\section{Discussion on Multi-Scale ECG Interpretation}
\label{sec:multi-level-design}
% \subsection{Examples of ECG Reports Containing Token-level Description.}
%
While some portion of ECG reports contain high-level summaries like ``sinus rhythm," many real-world diagnostic reports also include detailed references to waveform-level features. 
To further support this, we provide representative examples from our pretraining dataset, MIMIC-IV-ECG database ~\cite{gow2023mimic} in Table~\ref{tab:examples}.
These demonstrate that detailed morphological patterns at the waveform (token) level are frequently described in the reports.
\vspace{-0.4cm}
\begin{table}[ht!]
    \centering
    \caption{Examples of ECG reports from MIMIC-IV-ECG dataset. Each row is a complete text report and those parts in bold are descriptions about fine-grained ECG details.}
    \begin{tabular}{p{16cm}}
        \toprule
        Regular rhythm. Lead(s) unsuitable for analysis: \textbf{V1. Q waves in inferior leads}. \textbf{T wave inversion also present}. Possible inferior infarction – age undetermined. \textbf{Anterolateral ST-T changes}. Summary: abnormal ECG. \\
        \midrule
        Sinus tachycardia. \textbf{Short PR interval}. Borderline ECG. \\
        \midrule
        Sinus rhythm. \textbf{Poor R wave progression} – probable normal variant. \textbf{Anterolateral T wave changes} may be due to myocardial ischemia. Abnormal ECG. \\
        \midrule
        Atrial fibrillation. \textbf{Extensive ST-T changes are nonspecific}. Abnormal ECG. \\
        \midrule
        Probable accelerated junctional rhythm. \textbf{Low QRS voltages in limb leads}. Abnormal ECG. \\
        \bottomrule
    \end{tabular}
    \vspace{-0.4cm}
    \label{tab:examples}
\end{table}

Furthermore, even broader assessments like ``sinus rhythm" are based on a set of well-established low-level criteria, such as the presence of a P wave before every QRS complex, upright P waves in leads I, II, and aVF, regular R-R intervals, consistent PR intervals, and a normal heart rate. 
More examples can be found in Table~\ref{tab: cardio}.
Although these features are not explicitly mentioned in the cardiology reports in the MIMIC-IV-ECG dataset, these diagnosis are actually derived from these fine-grained waveform analysis and can be effectively captured by token-level representations during the learning process.
\vspace{-0.4cm}
\begin{table*}[ht!]
    \centering
    \caption{
    Cardiology examples}
    \label{tab: cardio}
    \resizebox{\textwidth}{!}{
    \begin{tabular}{l|l|l}
    \hline
    clinical diagnosis                                     & ecg criteria                                                                                  & whether contain local descriptions \\ \hline
    \multirow{3}{*}{Left Anterior Fascicular Block (LAFB)} & rS   complexes in leads II, III, aVF, with small R waves and deep S waves                     & True (Token-level)                 \\
                                                           & qR complexes in leads I, aVL, with   small Q waves and tall R waves                           & True (Token-level)                 \\
                                                           & Left Axis Deviation (LAD): Leads II, III and aVF are NEGATIVE;   Leads I and aVL are POSITIVE & False (rhythm-level)             \\ \hline
    \multirow{4}{*}{Atrial Fibrillation}                   & Irregularly irregular rhythm                                                                  & False (rhythm-level)              \\
                                                           & No P waves                                                                                    & True  (Token-level)              \\
                                                           & QRS   complexes usually \textless 120ms                                                       & True   (Token-level)              \\
                                                           & Variable ventricular rate                                                                     & False (rhythm-level)              \\ \hline
    \multirow{5}{*}{Left Bundle Branch Block (LBBB)}       & QRS duration $\geq$ 120ms                                                                     & True  (Token-level)              \\
                                                           & Dominant S wave in V1                                                                         & True  (Token-level)              \\
                                                           & Broad monophasic R wave in lateral leads (I, aVL, V5-6)                                       & True  (Token-level)              \\
                                                           & Absence of Q waves in lateral leads                                                           & True  (Token-level)               \\
                                                           & Prolonged R wave peak time \textgreater 60ms in leads V5-6                                    & True   (Token-level)              \\ \hline
    \multirow{5}{*}{sinus rhythm}                          & Regular   rhythm at a rate of 60-100 bpm                                                      & False  (rhythm-level)             \\
                                                           & Each QRS complex is preceded by a normal P wave                                               & True   (beat-level)               \\
                                                           & Normal P wave axis: P waves upright in leads I and II, inverted in aVR                        & True   (token-level)             \\
                                                           & The PR interval remains constant                                                              & True   (token-level)             \\
                                                           & QRS   complexes \textless 100 ms wide                                                         & True   (token-level)              \\ \hline
    \end{tabular}
}  
\end{table*}

\vspace{-0.4cm}
Building on these observations, we argue that using token-level ECG embeddings for report generation is not only appropriate but essential for capturing the full spectrum of clinically meaningful information. To clarify this motivation, we have revised the manuscript and included supporting examples that illustrate the rationale behind our design choice.

Moreover, we believe that incorporating more fine-grained textual descriptions—such as explicit references to waveform components—alongside general diagnostic terms could further strengthen our approach. These detailed references would provide richer supervision signals and more closely align with established clinical diagnostic criteria.
For instance, broad terms like “sinus rhythm” could be supplemented with explicit criteria such as the presence of P waves before each QRS complex and consistent PR intervals. Integrating such detail into the training data may improve alignment between ECG signals and textual descriptions. While this direction holds promise, it is beyond the scope of the current work and is left for future exploration.

\section{Additional Experimental results}
\subsection{Results for Patient Identification task}
\label{sec:patient_identification}
\begin{table}[ht!]
    \centering
    \caption{Zero-shot Patient identification results using Top-k recall [\%].
    Here we use 608 pairs for patient identification following~\citet{oh2022lead}.}
    \resizebox{0.6\textwidth}{!}{
    \begin{tabular}{c | c c c}
        \toprule
        Method & \multicolumn{3}{c}{PTBXL} \\
        & R@1 & R@5 & R@10 \\ 
        \midrule
        Wav2Vec 2.0 + CMSC + RLM~\cite{oh2022lead} & 39.8 & 52.14 & 59.21 \\
        ECGFM~\cite{mckeen2024ecg} & 49.18 & 60.70 & 67.76 \\
        MERL~\cite{liu2024merl} & 16.12 & 26.32 & 31.74 \\
        \modelname & \textbf{49.67} & \textbf{66.12} & \textbf{70.89} \\
        \bottomrule        
    \end{tabular}
    }
    \vspace{-0.4cm}
    \label{tab:patient_identification_results}
\end{table}
The patient identification task trains a model to recognize individual patients based on their ECG signals. It does this by learning to generate unique numerical representations  for each person's ECG. The core objective is to ensure ECGs from the same patient produce highly similar representations, while ECGs from different patients produce distinct ones. For detailed experimental procedures, we follow the settings of ~\cite{oh2022lead}. As shown in Table \ref{tab:patient_identification_results}, our model \modelname  outperforms all three baselines in top-1, top-5, and top-10 accuracy.
\vspace{-0.2cm}
\begin{table*}[ht!]
    \centering
    \caption{Distribution of number of beats in the training set of MIMIC-IV-ECG.}
    \label{tab:heart_beat_dis}
    \resizebox{0.9\linewidth}{!}{
    \begin{tabular}{c c c c c c c c c c c c c c c}
        \toprule
        Beat Count  & 8 & 9 & 10 & 11 & 12 & 13 & 14 & 15 & 16 & 17 & 18 & 19 & 20  & Others\\
        \midrule
        Frequency & 18357 & 47635 & 93075 & 112010 & 112424 & 93830 & 74027 & 62150 &  47997 & 24509 & 15987 & 11606 & 8347 & 23493\\
        \midrule
        Percentage & $2.5\%$& $6.4\%$ & $12.5\%$ & $15.0\%$ & $15.1\%$ & $12.6\%$ & $9.9\%$ & $8.3\%$ &  $6.4\%$ & $3.3\%$& $2.1\%$& $1.6\%$ & $1.1\%$& $3.2\%$\\
        \bottomrule
    \end{tabular}
    }
\end{table*}
\vspace{-0.4cm}
\vspace{-0.4cm}
\begin{table*}[ht!]
    \centering
    \caption{
    Ablation results of the number of learnable beats.}
    \label{tab:ablation_of_beats}
    \resizebox{0.95\textwidth}{!}{
        \begin{tabular}{ c | c c c c c c c c c c c c c c c c c c | c}
            \toprule
            \multirow{2}{*}{\#. Beats} 
             & \multicolumn{3}{c}{PTBXL-Rhythm}  & \multicolumn{3}{c}{PTBXL-Form} & \multicolumn{3}{c}{PTBXL-Sub} &  
            \multicolumn{3}{c}{PTBXL-Super} & \multicolumn{3}{c}{CPSC2018} & \multicolumn{3}{c}{CSN} & \multirow{2}{*}{Average}  \\
             & 1\% & 10\% & 100\% & 1\% & 10\% & 100\% & 1\% & 10\% & 100\% & 1\% & 10\% & 100\% & 1\% & 10\% & 100\% & 1\% & 10\% & 100\% \\
            \midrule 
            10 & \textit{88.83}    & 94.65          & 96.91          & 63.41          & 76.71          & 83.30          & 79.22          & 84.40          & \textbf{87.46} & \textbf{85.82} & \textit{87.61}    & \textit{87.87}    & 88.54          & 91.75          & \textbf{94.32} & 78.25          & 84.83          & 90.17          & 85.78          \\
            12 & 87.98          & 94.73          & 96.81          & 62.33          & 76.94          & 84.35          & 79.69          & 84.99          & 86.86          & 85.61          & 87.57          & 87.79          & \textit{88.58}    & 92.70          & 93.76          & \textbf{79.89} & 87.22          & 90.29          & 86.00          \\
            14 & 87.12          & 95.82          & 96.53          & \textit{64.11}    & \textbf{78.92} & \textbf{84.80} & 80.30          & \textbf{85.98} & \textit{87.31}    & 85.39          & 87.40          & 87.66          & 87.58          & \textit{92.84}    & \textit{94.14}    & \textit{79.11}    & \textbf{87.87} & \textbf{91.50} & \textit{86.35}    \\
            16 & \textbf{90.37} & \textbf{96.68} & \textbf{97.32} & \textbf{64.74} & 76.91          & 83.21          & \textbf{80.66} & \textit{85.17}    & 87.03          & 85.56          & 87.48          & 87.49          & \textbf{89.18} & \textbf{93.15} & 94.07          & 78.91          & 87.18          & 90.23          & \textbf{86.41} \\
            18 & 88.01          & \textit{96.48}    & \textit{97.20}    & 62.71          & \textit{77.51}    & \textit{83.98}    & \textit{80.37}    & 84.99          & 86.88          & \textit{85.68}    & \textbf{87.63} & \textbf{87.93} & 87.65          & 92.75          & 93.74          & 78.48          & \textit{87.23}    & \textit{91.25}    & 86.14 \\
            \bottomrule
        \end{tabular}
    }
    \vspace{-0.4cm}
\end{table*}
\subsection{Ablation on Number of Beats.}
\label{sec:heart_beats}
In the main text, we have chosen 10 heart beats is because we assume that the majority coherts in MIMIC-IV-ECG database have 10 heart beats for each ECG recordings. (The sample duration is 10 seconds and normal heart beat is 1 beat per second).
However, we have carefully calculate this statistics and found that median number of heart beats are 12 - 13, as evidenced by Table \ref{tab:heart_beat_dis}.
Thus, we have conducted further ablation study to explore more insights of this hyperparameter, and the results are as below in Table \ref{tab:ablation_of_beats}:
The model with 16 learnable heart beats performance best among all variants and have a performance gain of +0.66\% compared with 10 learnable beats setting.

\begin{table*}[t!]
    \centering
    \caption{Detailed statistics of the datasets.}
    \label{tab:dataset_stats}
    \resizebox{\linewidth}{!}{
    \begin{tabular}{l|cccccc|cccccccccc|ccccc}
    \hline
    \textbf{Split} & \multicolumn{6}{c|}{\textbf{PTB-XL}} & \multicolumn{9}{c|}{\textbf{CPSC2018}} & \multicolumn{4}{c}{\textbf{CSN}} \\
                  & NORM & CD & HYP & MI & STTC & Total & NORM & AF & I-AVB & LBBB & RBBB & PAC & PVG & STD & STE & Total & AF & GSVT & SB & SR & Total \\ \hline
    Train         & 7254 & 2048 & 1353 & 416 & 1907 & 12978 & 1213 & 1289 & 889 & 243 & 1964 & 864 & 1084 & 1148 & 264 & 8958 & 1583 & 1639 & 2804 & 1625 & 7651 \\
    Val           & 916 & 234 & 172 & 64 & 256 & 1642 & 197 & 168 & 101 & 33 & 292 & 130 & 146 & 178 & 58 & 1303 & 186 & 189 & 315 & 161 & 851 \\
    Test          & 913 & 256 & 184 & 56 & 243 & 1652 & 365 & 342 & 251 & 56 & 589 & 274 & 308 & 345 & 68 & 2598 & 449 & 472 & 769 & 436 & 2126 \\ \hline
    \end{tabular}
    }
    \vspace{-0.4cm}
\end{table*}

\section{Dataset Details.}
\label{sec:dataset_details}
Table~\ref{tab:dataset_stats} summarizes the key statistics of the datasets used in this study. The MIMIC-IV-ECG dataset provides a large corpus of 760,618 ECG records without explicit diagnostic labels, serving as unlabeled data for pretraining. In contrast, PTB-XL, CPSC2018, and CSN are labeled datasets used for downstream evaluation. PTB-XL comprises 12,978 training, 1,642 validation, and 1,652 test samples, spanning five diagnostic categories such as NORM, CD, and HYP. CPSC2018 includes 8,958 training, 1,303 validation, and 2,598 test samples across nine rhythm- and morphology-related classes, including AF, PAC, and RBBB. CSN consists of 7,651 training, 851 validation, and 2,126 test samples, with annotations for four diagnostic categories such as AF and GSVT. Together, these datasets encompass a broad spectrum of clinically relevant cardiac conditions, enabling robust assessment of model generalization across diverse domains.

\section{Implementation Details}  
\label{sec:pretraining_details}

\subsection{Evaluation Metric}  
\textit{AUC} (Area Under the Curve) quantifies the overall classification performance as the area under the ROC curve. Ranging between 0 and 1, a higher AUC indicates superior model discriminative power. Specifically, it estimates the probability that the model ranks a randomly selected positive instance higher than a negative instance. Our optimization objective prioritizes AUC maximization to achieve optimal TPR-FPR trade-offs.  
\begin{table*}[t!]
    \centering
    \caption{Domain transfer category matching.}
    \label{tab:transfer_description}
    \resizebox{0.7\textwidth}{!}{
    \begin{tabular}{l|llll}
        \toprule
        \textbf{Target Category} & \textbf{PTBXL-Super} & \textbf{CPSC2018} & \textbf{CSN} \\
        \midrule
        AFIB         & -         & AFIB       & AFIB \\
        APB          & -         & PAC        & - \\
        CLBBB        & CD        & -          & - \\
        CRBBB        & CD        & -          & - \\
        HYP          & HYP       & -          & RVH, LVH \\
        LBBB         & CD        & -          & CLBBB \\
        MI           & MI        & -          & MI \\
        NORM         & NORM      & NORM       & SR \\
        PAC          & -         & PAC        & - \\
        SR           & -         & NORM       & SR \\
        STD          & STTC      & STD        & STE, STTC, STTU, STDD \\
        STE          & STTC      & STE        & STE \\
        STTC         & STTC      & -          & STTC, STE, TWO, STTU, QTIE, TWC \\
        VPC / VPB    & -         & VPC        & VPB \\
        1AVB         & CD        & 1AVB       & 1AVB \\
        2AVB / AVB   & -         & -          & 2AVB, 2AVB1, AVB \\
        RBBB         & CD        & CRBBB      & RBBB \\
        \bottomrule
    \end{tabular}
    }
    \vspace{-0.6cm}
\end{table*}
\begin{table*}[ht!]
    \centering
    \caption{Network architecture of \modelname.}
    \label{tab:architecture}
    \resizebox{\linewidth}{!}{
    \begin{tabular}{l|c|l}
        \toprule
        \textbf{Block Name} & \textbf{Layer Number} & \textbf{Layer Components} \\
        \midrule
        ECG Transformer Encoder   & 8  & MultiHeadAttention, Dropout, LayerNorm, FC layer, LayerNorm \\
        ECG Feature Extractor     & 4  & Conv1d, Dropout, Fp32GroupNorm, GELU \\
        ECG Positional Encoding   & 1  & Conv1d, SamePad, GELU \\
        % ECG Quantizer             & 1  & GumbelVectorQuantizer, Linear \\
        ECG Attentional Pooler    & 1  & MultiheadAttention, LayerNorm \\
        Text Attentional Pooler   & 1  & MultiheadAttention, LayerNorm \\
        Text Encoder              & 12 & BertAttention, BertIntermediate, BertOutput \\
        Text Projector            & 1  & Linear, GELU, Linear \\
        Text Decoder              & 6  & ResidualAttentionBlock, LayerNorm, MultiheadAttention, Identity, LayerNorm, MLP, Identity \\
        \bottomrule
    \end{tabular}
    }
    \vspace{-0.4cm}
\end{table*}
\subsection{Domain Transfer Experimental Details}  
Table~\ref{tab:transfer_description} provides the label mappings used for domain transfer experiments.
We adopt the SCP-code label alignment protocol as proposed in~\cite{liu2024merl}.
Specifically, we conduct transfer learning by training on one dataset and evaluating on another, using the aligned target labels to assess generalization under domain shifts.
Since \modelname\ supports zero-shot classification, transfer learning is performed by directly evaluating its predictions on the overlapping label set between source and target datasets.
Categories without a direct correspondence are excluded from evaluation to ensure consistency and fairness in performance comparison.

\subsection{Network Architecture}
Table~\ref{tab:architecture} presents the detailed network architecture of \modelname, which comprises modules for both the ECG encoder and the text encoder.
The ECG encoder includes an 8-layer Transformer Encoder with multi-head attention, dropout, layer normalization, and feedforward layers, designed to model long-range temporal dependencies. It is preceded by an ECG Feature Extractor composed of 4 blocks, each containing Conv1d, dropout, GroupNorm, and GELU activation. Positional information is encoded using a dedicated Conv1d-based Positional Encoding block. To extract semantically meaningful pooled embeddings, both the ECG and text branches include an Attentional Pooler built with multi-head attention and layer normalization.

The text encoder consists of a 12-layer Transformer, followed by a lightweight Text Projector for dimensionality alignment. Notably, we implement causal attention in the text encoder to prevent information leakage. A 6-layer Text Decoder integrates residual attention blocks and multi-head attention mechanisms to produce cross-modal outputs.

\end{document}